\documentclass[10pt,a4paper]{report}
\usepackage[latin1]{inputenc}
\usepackage{amsmath,color}
\usepackage{booktabs}
\usepackage[margin=2.7cm]{geometry}
\usepackage{amsfonts}
\usepackage{amssymb}
\usepackage[colorlinks=true,urlcolor=blue,citecolor=brown]{hyperref}
\usepackage{tikz}
\usepackage{graphicx}
\author{Mohamad Shalaby (Mohamad@aims.ac.za)} 
\title{Dynamics and Light Propagation \\ in a Universe with Discrete Matter Content}
\begin{document}
\date{}
\maketitle 
\section*{Abstract} 

We discuss a model for a universe with discrete matter content instead of the continuous perfect fluid taken in FRW models. We show how the redshift in such a universe deviates from the corresponding one in an FRW cosmology. This illustrates the fact that averaging the matter content in a universe and then evolving it in time, is not the same as evolving a universe with discrete matter content. The main reason for such deviation is the fact that the photons in such a universe mainly travel in an empty space rather than the continuous perfect fluid in FRW geometry.
\tableofcontents
\chapter{Introduction}

The main goal for this work is to show a model for the universe in which it is not assumed that the matter content of the universe is, on average, a perfect continuous fluid which permeates all the universe. We discuss a more realistic model in which the matter content is assumed to be discrete and show how this does not affect the dynamics of the universe but affects the light propagation in the universe.

The main motivation behind discussing such model is that in FRW cosmology, the most usable model for the universe, the matter content is assumed to be a perfect continuous fluid which permeates the universe. To use this model to interpret the data, observed experimentally, we should assume the existence of an exotic continuous perfect fluid with a constant density through all the universe i.e.\@ it never dilutes as the universe expand. This perfect fluid is called dark energy. Therefore, considering such an inhomogeneous model could provide further insight to investigate the validity of assuming the existence of dark energy.

We start off in Chapter 2 by discussing Einstein's picture of gravity and the geometric consequences for this approach; we discuss the geometric description of the Einstein field equations in a general curved spacetime. We then discuss the general form of the energy--momentum tensor and then write it explicitly for the perfect fluid case in a general frame of reference.

In Chapter 3, we discuss two exact solutions for Einstein's field equations, namely, the Schwarzschild solution, and the Friedmann-Robertson-Walker (FRW) solution.\@ In each geometry, we discuss the equations of motion of photons and massive particles in the resulting spacetime and also the redshift.\footnote{The redshift is one of the most important quantities in observational cosmology.} In the FRW solution, we get two independent equations for the scale factor of the universe. We then discuss the three solutions in the case of the matter-dominated universe, namely open, closed and flat universes in that case.

In Chapter 4, we use this quick introduction to discuss a discrete model for the universe introduced by Lindquist and Wheeler (LW) in 1957 in a seminal paper \cite{lwmodel}. They discussed in their model the case of a matter-dominated closed universe and showed that the global dynamics are similar to the corresponding case in the FRW cosmology. We then discuss how this model was extended in another wonderful paper  by T.~Clifton and P.~Ferreira in 2009 \cite{discos} to discuss the light propagation for such a set-up. In the case of a flat universe, they showed a deviation in the redshift from the corresponding redshift in an FRW cosmology, namely, the Einstein--de--Sitter (EdS) model. The main advantage of such a model is that that global dynamics of the FRW cosmology was shown to be a limiting case of that model, namely, when the number of the discrete islands becomes very large.

It is also important for the reader to notice that in the whole of this work, we are taking the metric with a~$-2$ signature. Otherwise stated, the sign of the time part is positive and the sign of the spatial part is negative, i.e.\@ in flat spacetime the metric is expressed as $ds^2 = dt^2 - dx^2 - dy^2 - dz^2$.
\chapter{Einstein Theory for Gravity}

\section{A different approach to treat the gravitational force}

Two important observations were a cornerstone, and a main motivation for Einstein in developing the General Theory of Relativity, a theory for gravity. First, the experimental fact that the inertial mass\footnote{Inertial mass represents the ability of a massive particle to resist any change in its state of motion.} of a massive particle is the same as its gravitational mass.\footnote{Gravitational mass represents the ability of a massive particle to accelerate toward other fixed gravitating object.} And second, the Equivalence Principle which says that in a freely falling and non-rotating part of space, i.e.\@ an elevator, the laws of physics are the same as the special relativity laws. Otherwise stated, if we consider only a uniform gravitational field in a closed elevator, i.e\@ we ignore the tidal forces, we can't distinguish between the gravitational field and the field which would be developed due to a uniform acceleration for that part of space. This means that all the particles in that region of space have no acceleration in the inertial frame of reference moving with this region of space.

This led Einstein to think of gravity as a curvature in the spacetime which is caused by matter and not as a force. Einstein's main idea was to replace the equation of motion for a particle moving under gravitational forces by the equation of motion for a freely-moving particle in a curved space in which the curvature depends on the amount of matter and energy in it, i.e.
\[\frac{d\textbf{P}}{d\tau} = 0.\]
Here, $\textbf{P}$ is the four-momentum, and $\tau$ is the proper time.

This means that the spacetime trajectory of the freely-moving particle in curved spacetime is a geodesic of that spacetime.


\section{Einstein field equations of gravity}

They are a set of 10-independent, non-linear differential equations for the metric fields $g_{\mu \nu}(x^{\mu})$,\footnote{Throughout this essay we consider $\mu, \nu,\delta,\lambda ,\rho ,\sigma, a, b, c, d, e =  0,1,2,3$ (spacetime), but $ i,j,k,l = 1,2,3$ (space only).} they relate the spacetime curvature to all the sources of energy in it, i.e.\@ ordinary matter, radiation, and dark energy. They can be written in tensorial form as follows.
\begin{itemize}
\item With a zero cosmological constant, $ \Lambda$, i.e.\@ the amount of the dark energy is zero,
\begin{equation} \label{wocc}
R_{\mu \nu} - \frac{1}{2} R g_{\mu \nu} = -\kappa T_{\mu \nu}.
\end{equation}
\item With a non-zero cosmological constant, $ \Lambda$,
\begin{equation} \label{wcc}
R_{\mu \nu} - \frac{1}{2} R g_{\mu \nu} = -\kappa T_{\mu \nu} - \Lambda g_{\mu \nu}. 
\end{equation}
\end{itemize}
Here, $R_{\mu \nu}$ is the Ricci tensor, $g_{\mu \nu}$ is the metric tensor, $R$ is the curvature scalar, and $\kappa = \dfrac{8 \pi G}{3 c^2}$ with the Newtonian gravitational constant $G$.

It is common to call the right-hand side of Einstein field equations as the Einstein tensor $G_{\mu\nu} = R_{\mu \nu} - ~\frac{1}{2} R g_{\mu \nu}$. 

Also, $ R_{ab} $ and $g_{ab} $ are related by  
\begin{equation} \label{riccit}
R_{ab} = {R^c}_{ ab c } = {\Gamma^c}_{k b} {\Gamma^k}_{ac} -{\Gamma^c}_{k c} {\Gamma^k}_{a b} + \partial_b {\Gamma^c}_{a c} -\partial_c {\Gamma^c}_{a b},
\end{equation}
where $ {R^d}_{ abc } $ is the general curvature tensor, or the Riemann-Christoffel tensor, and $ {\Gamma^b}_{a c} $ is the affine connection.

The affine connection is related to the metric elements by 
\begin{equation} \label{affine}
{\Gamma^a}_{b c}  = \frac{1}{2} g^{ad} (\partial_b g_{dc} + \partial_c g_{db} - \partial_d g_{bc} ).
\end{equation} 
Equations \eqref{riccit} and \eqref{affine} say that the Ricci tensor components depend on the metric components $g_{\mu \nu}$ and their derivatives. Therefore, we measure the spacetime curvature\footnote{The spacetime deviation from being flat.} using $g_{\mu \nu}$. This is the reason for considering the field equations as a differential equations on the metric fields $g_{\mu \nu}$.

If we contract equations \eqref{wocc} and \eqref{wcc} with $ g_{\mu \nu} $, then,
\begin{itemize}
\item from equation \eqref{wocc}, we get $ R = \kappa T$, therefore the Einstein field equations, with zero cosmological constant, can be rewritten as
\begin{equation} \label{wwcc}
R_{\mu \nu} = -\kappa ( T_{\mu \nu} - \frac{1}{2} T g_{\mu \nu} ).
\end{equation}
\item from equation \eqref{wcc}, we get $R = \kappa T + 4 \Lambda$, therefore the Einstein field equations, with non-zero cosmological constant, can be rewritten as
\begin{equation} \label{wwocc}
R_{\mu \nu} = -\kappa ( T_{\mu \nu} - \frac{1}{2} T g_{\mu \nu} ) + \Lambda g_{\mu \nu}.
\end{equation}
\end{itemize}
Here, $ T = g_{\mu \nu} T^{\mu \nu} = T ^{  \mu}_{\mu}$.

Now we can write equations equivalent to the equations of motion of particles moving under the influence of a gravitational field. These will be the equations of the geodesic\footnote{Geodesic line is the shortest possible line between two points in a general curved spacetime.} in a general curved spacetime, if that spacetime is curved due to the sources of that gravitational field. If the general trajectory in the spacetime is $x^{\mu}(s)$, where $s$ is the parameterization parameter along that curve, the tangent for that curve will be\footnote{Here, the dot indicates the derivative with respect to $s$.}
\[u^{\mu} = \dot{x^{\mu}} = \frac{dx^{\mu}}{ds},\]
and for that curve to be a geodesic, the covariant derivative of $\dot{x^{\mu}}$ should vanish, so we write the geodesic equations as\footnote{For a more detailed mathematical description the reader is referred to \cite{hobson}.}
\begin{equation} \label{geodesic}
\nabla_{\mu} u^{\mu} = \dot{ u^{\mu} } + \Gamma^{\mu}_{\nu \sigma } u^{\nu} u^{\sigma} = \frac{ d^2 x^{\mu} }{ds^2} + {\Gamma^\mu}_{\nu \sigma } \frac{dx^{\nu}}{ds} \frac{dx^{\sigma}}{ds} = 0.
\end{equation}
The full trajectory for the particle, i.e.\@ the solution of the equations of motion, is then given by the functions $x^{\mu}(s),$ that obey equation \eqref{geodesic}.

Provided that we accept the new approach for gravity, we can conclude from the above discussion that the metric components of that curved spacetime play an essential role in determining the trajectories of particles moving under the effect of gravity. They depend only on the energies in the spacetime which cause the spacetime to curve through the Einstein field equations. 
\section{Description of the Einstein field equations}
\subsection{ The spacetime curvature-sources side of the equation} 
The right-hand side of the equation \eqref{wcc} contains two different kinds of energy: the energy due to matter and radiation, in $T_{\mu \nu}$; and the cosmological constant, which expresses the dark/vacuum energy, i.e. the energy in the universe if we remove all the matter and the radiation from it.\footnote{In other words, one can think of the dark energy as the ground state of a quantized-bounded system, which is normally not of zero energy because of the Heisenberg uncertainty principle.}

The energy-momentum tensor, $T^{\mu \nu}$,\footnote{$T^{\mu \nu}$ is called sometimes the stress-energy tensor} is the term which expresses the energy and the momentum flow for radiation and matter, as a source of the gravitational field that curves the spacetime, in general it can be expressed as
\begin{equation} \label{Tmn}
T^{\mu \nu}(x^{\mu}) = \rho(x^{\mu}) u^{\mu} u^{\nu}.
\end{equation}
Here, $u^{\mu}=u^{\mu}(x^{\mu}) = \gamma_{u} (c,\vec{u}(x^{\mu}) )$ and $\rho$ is the energy density.

All the above quantities, $ \rho$, $ T^{\mu \nu}$, and $ u^{\mu}$ are measured in the frame of reference described by the coordinate system $x^{\mu}$ and at a certain point in that spacetime. Therefore, at that point  
\begin{itemize}
\item $T^{00} = \gamma^2_u \rho c^2$ expresses the energy density of the gravitating particle, 
\item $T^{0i}= \gamma^2_u \rho c u^{i}$ relates to the energy flux in the $i^{th}$ direction of the space,
\item $T^{i0}= \gamma^2_u \rho c u^{i} $ relates to the momentum flux in the $i^{th}$ direction of the space, and
\item $T^{ij}= \gamma^2_u \rho u^{i}u^{j}$ represents the rate of the flow of the $i^{th}$ component of the momentum in the $j^{th}$ direction.
\end{itemize}
Conservation of the energy and momentum can be expressed in the general spacetime as
\begin{equation} \label{conservation}
\nabla_{\mu} T^{\mu \nu  } = \partial_{\mu} T^{\mu \nu } + \Gamma^{\mu}_{\mu \lambda } T^{\lambda \nu }  = 0,
\end{equation}
where $\nabla_{\mu}$ is the covariant derivative operator.
\subsection{ The Spacetime curvature side of the equation} 

The left-hand side of equation \eqref{wocc}, or \eqref{wcc}, measures how much the spacetime is curved or deformed from being flat due to the existence of the deformation sources, i.e. any type of energy mentioned above. The Ricci tensor measures how much the spacetime, a 4-dimensional manifold, deviates from being a 4-dimensional Euclidean space. As indicated in the previous section, the determining quantities on this side of the equation are the metric tensor quantities, $g_{\mu \nu}$.

In the literature, there are many routes to derive the Einstein field equations. One of them, as mentioned before, is to use the conservation of energy condition \eqref{conservation} and the fact that these equations should, in non-relativistic limit, be reduced to the Newtonian field equations, to restrict the possible forms of the equations, for more details about the derivation the reader is referred to \cite{hobson},or to \cite{carroll}.

Another way to derive them is to take the Einstein-Hilbert action $ S = \int_{ \mathcal{R} } R \sqrt{\vert g \vert } d^4 x$, and apply the variational principle, the principle of least action, to get the field equations. For more mathematical details, the reader is referred to \cite{geometry} section 37.4.
%
%
\section{The energy momentum tensor in some special cases} \label{cases}
\subsection{For a free space}
The energy-momentum tensor components for that part of spacetime vanish since it contains no matter, i.e.\@ $ T^{\mu \nu}= T = 0$. Assuming no dark energy, this gives, from equation \eqref{wwcc}, that $ R_{\mu \nu } = 0$. In presence of dark energy, this gives, from equation \eqref{wwocc}, that $R_{\mu \nu }= \Lambda g_{\mu \nu}.$
\subsection{For a perfect fluid}
The advantage of taking the perfect fluid case is that in this approximation there is no interaction between the particles, i.e.\@ the sheer forces are ignored, and the thermal motion of the particles can be ignored. Using these assumptions, we can write $T$ for a perfect fluid in its instantaneous frame of reference (IFR) as
\begin{equation}
T = [ T^{\mu \nu} ]= 
\begin{bmatrix}
c^2 \rho & 0 & 0 & 0 \\ 0 & p & 0 & 0 \\0 & 0 & p & 0 \\0 & 0 & 0 & p
\end{bmatrix},
\end{equation}
where $p$ is the pressure of the fluid, and $\rho$ is the energy density of the fluid. It can also be written in tensorial form as 
\[T^{\mu \nu} = (\rho + p/c^2) u^{\mu} u^{\nu} - p \eta^{\mu \nu},\]
where $\eta^{\mu \nu}$ is the Minkowski metric which is given by 
\[\eta^{\mu \nu} = \begin{bmatrix}
1 & 0 & 0 & 0 \\ 0 & -1 & 0 & 0 \\0 & 0 & -1 & 0 \\0 & 0 & 0 & -1 
\end{bmatrix}
.\]
Therefore, the full tensorial expression for $T$ in any arbitrary coordinate system is
\begin{equation} \label{Tperfect}
T^{\mu \nu} = (\rho + p/c^2) u^{\mu} u^{\nu} - p g^{\mu \nu}.
\end{equation}
If we substitute this $T^{\mu \nu}$, given in equation \eqref{Tperfect}, in the covariant expression of the conservation of energy, given by equation \eqref{conservation}, we get the relativistic continuity equation and the relativistic equation of motion of a perfect fluid. Namely, in flat spacetime it becomes $ \partial_{\mu} T^{\mu \nu} = 0, $ which gives, using $u^{\mu} u_{\mu}=c^2$,
\begin{itemize}
\item the continuity equation
\begin{equation} \label{cont}
\partial_{\mu}(\rho u^{\mu}) + (p/c^2) \partial_{\mu} u^{\mu} = 0;
\end{equation}
\item and the equations of motion
\begin{equation}
(\rho + p/c^2) (  \partial_{\mu} u^{\nu} )u^{\mu} = (\eta^{\mu \nu} - u^{\mu}u^{\nu}/c^2) \partial_{\mu}p.
\end{equation}
\end{itemize}
\chapter{Exact Solutions for Einstein Field Equations}
\section{Introduction}
Accepting Einstein's new approach for treating the gravitational forces, we present here two exact solutions for his field equations. Namely, the Schwarzschild solution,\footnote{Schwarzschild was in the trenches on the Eastern Front conflict when he came up with his solution, and sadly he did not survive the conflict.} which is the spacetime geometry outside a spherically symmetric matter distribution, and the Friedmann-Robertson-Walker (FRW) solution, which is the spacetime geometry of a homogeneous and isotropic space containing a uniform distribution of matter.
%
%
%
\section{Schwarzschild geometry}
Since the space outside that spherically symmetric mass distribution is free,\footnote{It is important here for the reader to notice that we shall ignore the dark energy term in our treatment.} therefore, as we mentioned in Section \ref{cases},
\begin{equation} \label{aa}
R_{\mu \nu} = 0.
\end{equation}
The general form of a static, spherically symmetric metric in polar-spatial coordinates is 
\begin{equation} \label{112244}
ds^2 = A(r) dt^2 - B(r) dr^2 - r^2 d \theta ^2 - r^2 \sin^2 \theta d \phi ^2.
\end{equation}
Using condition \eqref{aa} and the line element \eqref{112244} we can solve for $A(r)$, and $B(r)$ and we then get the Schwarzschild line element,\footnote{For a detailed derivation the reader is referred to \cite{lawden} pages 142--7.} written as
\begin{equation} \label{Sch}
ds^2 = c^2 (1-2 m/r) dt^2 - \frac{ dr^2 }{(1-2 m/r)}- r^2 d \theta ^2 - r^2 \sin^2 \theta d \phi ^2.
\end{equation}
Here, $m = MG/c^2$, where $M$ is the total mass of the spherically symmetric mass distribution, $r$ is the Schwarzschild radial component,\footnote{$r = 2m$ is dubbed the Schwarzschild radius, and if the massive object contract within $r = 2m$, this object becomes Schwarzschild blackhole.} $t$ is the Schwarzschild time, and $\theta$ and $\phi$ are the polar angles.

By looking to the second term in equation \eqref{Sch}, we can see that $r$ is not the same as the radial distance in normal spherical polar coordinates. Also, from this expression for the metric we can interpret $t$ as the proper time experienced by a stationary observer at $r\rightarrow \infty $.

It was shown later by Birkhoff that the spherically symmetric vacuum solution is the same for a static and non--static spacetime\footnote{Here, we taking about the spacetime exterior to the matter distribution.}, since both have the same line element (Birkhoff theorem).

In the vicinity of this mass distribution $M$, following Einstein's proposal, the spacetime will be curved, and to determine how other objects, photons or massive particles, behave, we should determine the geodesic equations for a freely-moving object in a spacetime with the line element given by equation \eqref{Sch}.
\subsection{The red-shift in the Schwarzschild geometry}
For the case where both the emitter (E), and the receiver (R) are stationary in the space,\footnote{Stationary here means that $r$, $\theta$ and $\phi$ are of constant values.} the red-shift $z$ is expressed as
\begin{equation} \label{redshift}
1 + z = \frac{\nu _{E}}{ \nu _{R} } = \sqrt{ \frac{g_{00}(R)}{g_{00}(E)} } = \sqrt{ \frac{1-2m/r_R}{1-2m/r_E} },
\end{equation}
where $r_E$ and $r_R$ are the Schwarzschild radial coordinates for the emitter and the receiver, respectively.
For a detailed proof of equation \eqref{redshift}, the reader is referred to read the elegant proof in section 4.3 of \cite{foster}.
%
%
\subsection{The geodesic equations in the Schwarzschild geometry}
We derive the geodesic equations using the Euler--Lagrange method.\footnote{The following treatment is based on the treatment on \cite{hobson}.} The Lagrangian $L$ is given by
\[L = g_{\mu \nu} \dot{x}^{\mu} \dot{x}^{\nu} = c^2(1-2m/r)\dot{t}^2 - \frac{\dot{r}^2 }{1-2m/r} - r^2( \dot{\theta}^2 +  \sin^2 \theta \dot{ \phi} ^2 ). \]
If we apply the Euler--Lagrange equation, $ \dfrac{d}{ds} \left( \dfrac{dL}{\partial \dot{x}^{\mu}} \right)  = \dfrac{\partial L}{\partial x ^{\mu}}$, we get
\begin{equation} \label{a}
(1-2m/r) \dot{t} = k,
\end{equation}
\begin{equation} \label{b}
\frac{\ddot{r}}{1-2m/r} + \frac{ m c^2}{r^2} \dot{t}^2 -\frac{m}{r^2 (1-2m/r)^2} \dot{r}^2  - r ( \dot{\theta}^2 +  \sin ^2 \theta \dot{ \phi} ^2 ) = 0,
\end{equation}
\begin{equation} \label{c}
\ddot{\theta } + \frac{2}{r} \dot{r} \dot{\theta } - \sin \theta \cos \theta \dot{\phi}^2 = 0, 
\end{equation}
\begin{equation} \label{d}
r^2 \sin^2 \theta \dot{\phi} = h,
\end{equation}
where $k,$ and $h$ are constants. The moving particle total energy is $E = p_{\mu} u ^{\mu}$, in particular, for a particle at rest at infinite, i.e.\@ $\dot{r} = 0$, and $r = \infty $, its 4-velocity will be $u ^{\mu} = (1,0,0,0)$. Therefore,
\[E  = p_0 = g_{00} m_0 \dot{t} = k m_0 c^2 \Rightarrow k = \frac{E}{m_0 c^2}, \]
where $m_0$ is the rest mass of the moving particle. Directly from equation \eqref{d} we see that $h = -l$, where $l$ is the angular momentum $l$ of the particle.\footnote{$l = g_{33} \dot{\phi}  = -r^2  \sin^2 \theta \dot{\phi} =-h $.}

We then consider the particles moving in an equatorial plane\footnote{This idea of taking the equatorial plane will only simplify the calculations, with no loss of generality since it is always possible to rotate the spatial part of the Schwarzschild coordinates such that the trajectory of the particle lies in the $(r,\phi)$ plane.}, i.e.\@ we keep only solutions with $\theta = \pi /2$. This satisfies equation \eqref{c}, and simplifies the equations into
\begin{equation} \label{aaa}
(1-2m/r) \dot{t} = k ,
\end{equation}
\begin{equation} \label{bb}
\frac{\ddot{r}}{1-2m/r} + \frac{ m c^2}{r^2} \dot{t}^2 -\frac{m}{r^2 (1-2m/r)^2} \dot{r}^2  - r  \dot{ \phi} ^2= 0 ,
\end{equation}
\begin{equation} \label{dd}
r^2 \dot{\phi} = h.
\end{equation}

Also, we can simplify these equations further by considering one of the following constraints:
\begin{itemize}
\item $ g_{\mu \nu} \dot{x}^{\mu} \dot{x}^{\nu} = c^2$ for massive particles.
\item $ g_{\mu \nu} \dot{x}^{\mu} \dot{x}^{\nu} = 0$ for photons.
\end{itemize}
We then replace the second--order differential equation  \eqref{bb} by one of these constraints, a first--order differential equation. Now we have the geodesic equations, and by solving them, we get the functions $x^{\mu}(s)$ that represent the trajectory for the particle. Here we separate the two cases: massive particles, and photons.
\begin{enumerate}
\item \textbf{The trajectories for massive particles.}

Replacing equation \eqref{bb} by the condition $ g_{\mu \nu} \dot{x}^{\mu} \dot{x}^{\nu} = c^2$, we get 
\begin{equation} \label{m1}
(1-2m/r) \dot{t} = k, 
\end{equation}
\begin{equation} \label{m2}
 c^2(1-2m/r)\dot{t}^2 - \frac{\dot{r}^2 }{1-2m/r} - r^2   \dot{ \phi} ^2  = c^2,
\end{equation}
\begin{equation} \label{m3}
r^2 \dot{\phi} = h.
\end{equation}
By substituting equations \eqref{m3} and \eqref{m1} into equation \eqref{m2}, we get the energy equation
\begin{equation} \label{me}
\dot{r}^2 + \frac{h^2}{r^2} (1-2m/r) - \frac{2mc^2}{r} = c^2 (k^2 - 1).
\end{equation}
Using equation \eqref{me} and the fact that $\dfrac{dr}{d \tau } = \dfrac{h}{r^2} \dfrac{dr}{d \phi}$, with $u=1/r$, we get the $u$-equation, which determines the shape of the massive particles' orbits,
\begin{equation} \label{mu}
\frac{d^2 u}{d \phi ^2} + u = \frac{GM}{h^2} + \frac{3GM}{c^2} u^2.
\end{equation}
To investigate these two equations \eqref{mu} and \eqref{me} more, it is useful to consider two cases: radial motion and circular motion of the massive particles. 
\begin{itemize}
\item Massive particles moving radially.

We have  $\phi = $ constant $\Rightarrow \dot{\phi} = 0 \Rightarrow h = 0$, and hence, using $MG=mc^2$, the energy equation becomes
\[\dot{r}^2 = c^2(k^2-1) + \frac{2MG}{r} \Longrightarrow \ddot{r} = - \frac{MG}{r^2}.\]
For a particle dropped from rest at $r= R \Rightarrow c^2(k^2-1) = - \frac{2MG}{R} \Rightarrow \dot{r}^2 = 2MG \left( \frac{1}{r} - \frac{1}{R} \right) $.

If the particle is dropped from rest at infinity, i.e.\@ $R = \infty$, we get that $ \dot{t} = \frac{1}{1-2m/r} $ and, 
\begin{enumerate}
\item for the in-falling particles $ \dot{r} = - \sqrt{\frac{2MG}{r}},  $ and
\item for the out-falling particles $ \dot{r} =  \sqrt{\frac{2MG}{r}} . $
\end{enumerate}
For the in-falling case, we have $\frac{dr}{d \tau } = - \sqrt{\frac{2MG}{r}}$, which by integration gives 
\begin{equation} \label{111}
\tau = \frac{2}{3} \sqrt{\frac{1}{2mc^2}} \left( \sqrt{r_0 ^3} -  \sqrt{r ^3} \right),
\end{equation}
where $r_0$ is the Schwarzschild radial coordinate at $\tau = 0$.
To find the trajectories in the $(r,t)$-plane, we use $\frac{dr}{dt} = \frac{\dot{r}}{ \dot{t} } = - (1-2m/r) \sqrt{2mc^2/r}$, which by integration gives
\begin{eqnarray*}  
t &=& \frac{2}{3} \left( \sqrt{\frac{r_0^3}{2mc^2}} - \sqrt{\frac{r^3}{2mc^2}} \right) + \frac{4m}{c} \left( \sqrt{\frac{r_0}{2m}} - \sqrt{\frac{r}{2m}} \right)  + \\ && \frac{2m}{c} \ln \left \vert \left(    \frac{\sqrt{r/2m} +1 }{\sqrt{r/2m} -1}      \right) \left(      \frac{\sqrt{r_0/2m} -1 }{\sqrt{r_0/2m} +1}      \right)  \right \vert .
\end{eqnarray*}
We notice from the above two equations that at $r = 0$, $\tau$ is finite, but, at $r = 2m>0$, $t$  goes to infinity, which means that the particle takes finite proper time to reach $r=0$, but for a stationary observer at $r = \infty$, it takes infinite time  before reaching $r = 2m$. Figure \ref{sssf1} shows the radially in-falling trajectory of a particle released from rest at infinity.
\begin{figure}[h]
\centering
\includegraphics[width=9cm,height=9cm]{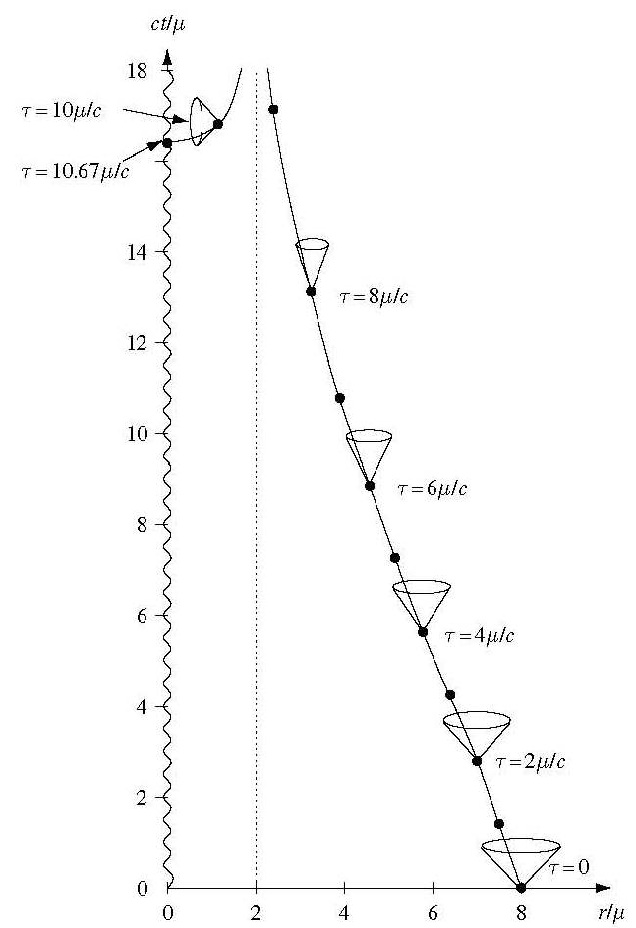}
\caption{Trajectory of a radially in-falling particle released from rest. The dots correspond to unit intervals of $c \tau /  m$, where $\tau$ is the particle's proper time and we have taken $ \tau = t = 0$ at $r_0 = 8m$.\protect\footnotemark}
\label{sssf1}
\end{figure}
\footnotetext{This figure is taken from \cite{hobson}, and in the figure $\mu$ is the same as our $m.$}
\item Massive particles moving in a circular orbit.

Here we have that $r= \mathrm{constant}$, and so $\dot{r}=\ddot{r} = 0 $, also $u= \mathrm{constant}$ or $\dfrac{d^2 u}{d \phi ^2} = 0$, consequently, the $u$-equation gives
\begin{equation} \label{mcu}
h^2 = \frac{MGr^2}{r-3m},
\end{equation}
and the energy equation gives
\begin{equation} \label{mce}
k = \frac{E}{m_0 c^2}=\frac{1-2m/r}{\sqrt{1-3m/r} }.
\end{equation}
Equation \eqref{mce} shows that the particles' orbits with radial coordinate $r \geq 4m$ are bounded, i.e.\@ $k<1$, the orbits with $3m<r<4m$ are not bounded, and for a particle to orbit at $r=3m$ it needs to have infinite energy. This is shown in Figure \ref{figure122}.
\begin{figure} [h]
\centering
\includegraphics[scale=0.85]{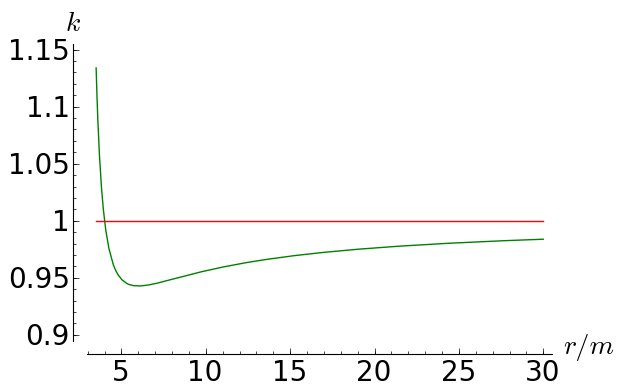}
\caption{This plot shows the dependence of $k  = E_{_{\text{total}}}/m_0 c^2$ on the radial component for that orbit. It shows that for a particle to be orbiting at $r=3m$, its total energy should be infinite, and the innermost stable orbit, i.e.\@ with the lowest energy, is at $r = 6m$.}
\label{figure122}
\end{figure}
\end{itemize}
\item \textbf{The trajectories for photons.}

Replacing equation \eqref{bb} by the condition $ g_{\mu \nu} \dot{x}^{\mu} \dot{x}^{\nu} = 0$, we get\footnote{For photons trajectories we can't use the proper time, $\tau$, as a parameterization parameter, so we use an arbitrary parameter $s.$}
\begin{equation} \label{p1}
(1-2m/r) \dot{t} = k ,
\end{equation}
\begin{equation} \label{p2}
 c^2(1-2m/r)\dot{t}^2 - \frac{\dot{r}^2 }{1-2m/r} - r^2  \dot{ \phi} ^2  = 0,
\end{equation}
\begin{equation} \label{p3}
r^2 \dot{\phi} = h.
\end{equation}
By substituting equations \eqref{p3} and \eqref{p1} into equation \eqref{p2}, we get the energy equation
\begin{equation} \label{pe}
\dot{r}^2 + \frac{h}{r^2} (1-2m/r) - \frac{2mc^2}{r} = c^2 k^2 .
\end{equation}
As for the massive particles case, using equation \eqref{me} and the fact that $\dfrac{dr}{d \tau } = \dfrac{h}{r^2} \dfrac{dr}{d \phi}$, we can write the $u$-equation as 
\begin{equation} \label{pu}
\frac{d^2 u}{d \phi ^2} + u = \frac{3GM}{c^2} u^2,
\end{equation}

To investigate these two equations \eqref{pu} and \eqref{pe} more, it is useful to consider the two cases of radial motion and circular motion of the photons. 
\begin{itemize}
\item Photos moving radially.

We have $\phi = $ constant $\Rightarrow \dot{\phi} = 0 \Rightarrow h = 0$, hence, the energy equation becomes
\begin{equation} \label{pr}
\dot{r} = \pm ck.\footnote{Here the + and the - signs correspond to the out-falling and the in-falling photos, respectively.} 
\end{equation}
Equation \eqref{pr} determines the trajectory in the $(r,s)$-plane, using it, with equation \eqref{p1}, we can write
\begin{equation} \label{prr}
\frac{dr}{dt} = \frac{\dot{r} }{\dot{t} } = \pm c (1-2m/r).
\end{equation}
At $r \rightarrow \infty$, from equation \eqref{prr}, we can write that $ \frac{dr}{dt} = \pm c $, which represents the normal light cone with slope $\pm 1$.

From equations \eqref{ofp}, and \eqref{ifp}, when $r \rightarrow 2m \Rightarrow$ the slope of the light cone~$\rightarrow~\pm~\infty$, which means that the light cone closes up when the Schwarzschild radial coordinate approaches $2m$, and that explains the result that the massive particles take infinite Schwarzschild time to approach the Schwarzschild radius $r=2m$. 
Figure \ref{sssf2} shows the structure of the lightcone in Schwarzschild's geometry, and how they close up when $r$ approachs $2m$.
\begin{figure}
\centering
\includegraphics[scale=1.0]{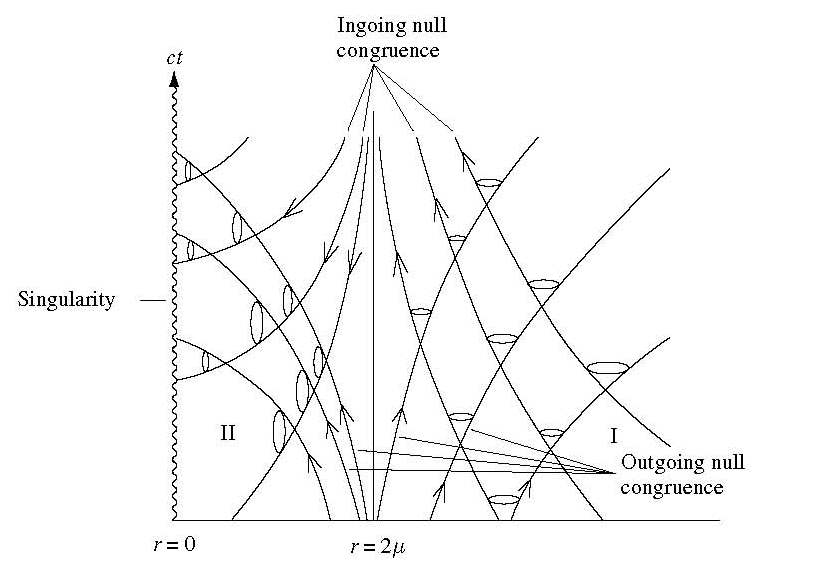}
\caption{Lightcone structure of the Schwarzschild solution.\protect\footnotemark}
\label{sssf2}
\end{figure}
\footnotetext{This figure is taken from \cite{hobson}, and in the figure $\mu$ is the same as our $m.$} 

Using equation \eqref{prr} we can learn about the trajectory in the $(r,t)$-plane, namely, 
\begin{itemize}
\item for in-falling photons,
\begin{equation} \label{ifp}
 ct = -r -2m \ln \vert r/2m -1 \vert  + \text{constant}.
\end{equation}
\item for out-falling photons,
\begin{equation} \label{ofp}
 c t = +r +2m \ln \vert r/2m -1 \vert  + \text{constant}.
\end{equation}
\end{itemize}
\item Photons moving in a circular orbits.

Here we have that $r=$constant $\Rightarrow \dot{r}=\ddot{r} = 0 $, and $u=$ constant $\Rightarrow \dfrac{d^2 u}{d \phi ^2} = 0$. Consequently, the $u$-equation gives that the only possible circular photons' orbit is at $r = \frac{3MG}{c^2}.$

For the sun, $\frac{3MG}{c^2} \simeq 4.5 $ km, but the sun radius is much larger than that, so we can't see such circular orbit for photons around the sun. However, in most cases, that is possible for black-holes.
\end{itemize}
\end{enumerate}
If we calculate the Riemann tensor components, $R_{ijkl},$ for the Schwarzschild line element \eqref{Sch}, we find some non-zero components which depend only on the Schwarzschild radial coordinate $r$. This shows that the spacetime geometry around a spherical mass distribution $M$ is not uniformly curved, but still spherically symmetric since these non-zero components don't depend on the polar angles.
%
%
\section{Friedmann-Robertson-Walker geometry}
The main reason for studying the geometry of isotropic and homogeneous space is that, on a large scale, the universe looks isotropic, which is also bolstered by the constancy of the cosmic microwave background radiation temperature. It is also believed that there is no preferred center for the universe, i.e.\@ there is no spacial place in the universe. This implies that the universe is also homogeneous.
\subsection{The metric for an isotropic and homogeneous space}
To describe a metric, contains such properties, we slice the spacetime into a continuous series of 3D spacelike\footnote{In general, in a spacelike hyper--surface any 2 events are separated by a spacelike distance in the spacetime, so we can always find an inertial frame of reference in which the 2 events occur at the same time, which is so called the surface of simultaneity.} hyper--surfaces.\footnote{In general, hyper--surface refers to an $n$-d manifold embedded in $n+1$ Euclidean space.}

\begin{figure}[here]
\centering
\includegraphics[scale=1]{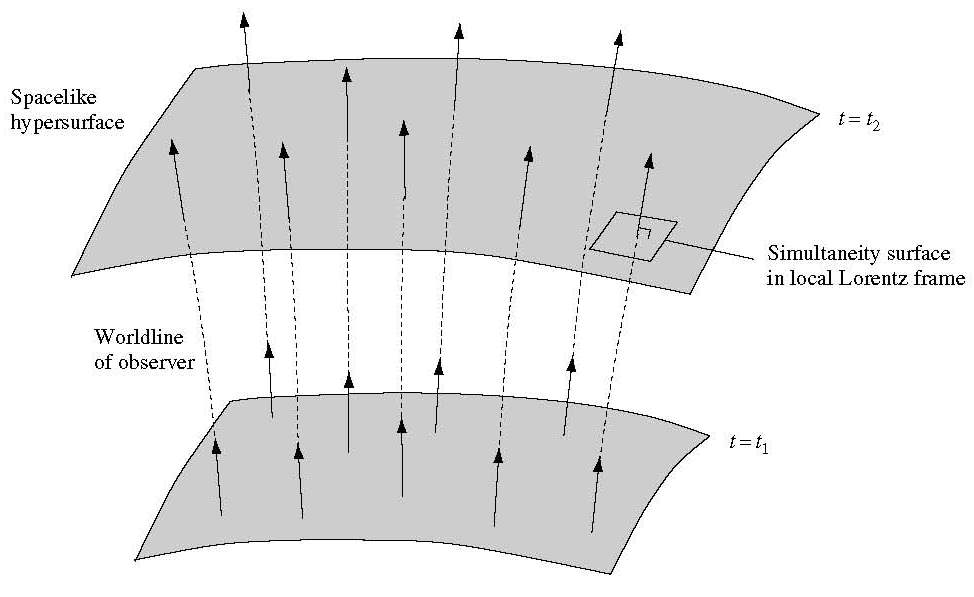}
\caption{Hyper--surfaces of simultaneity, to which the worldlines are orthogonal.\protect\footnotemark}
\label{figure1}
\end{figure}

To define global time, we slice the spacetime such that the hyper--surfaces form a non-intersecting spacelike continuous series, i.e.\@ a constant value of the parameter $t$ is assigned for each hyper--surface. Normally we construct these series, which can be constructed in many ways, such that the worldline for a comoving\footnote{Comoving usually means that as time evolves for that observer its spatial coordinate values, ($r,\theta,\phi$), remain constant.} observer is perpendicular to the hyper--surface, which the observer belongs to. This is shown in Figure \ref{figure1}.

For such slicing of spacetime, the worldlines are orthogonal to hyper--surfaces, which represent the spatial part of the spacetime. We can write the line element as\footnote{For more details about the following derivations the reader is referred to \cite{islam}, third chapter.}
\begin{equation} \label{l1}
ds^2=c^2 dt^2 - g_{ij}(t,x^1,x^2,x^3) dx^i dx^j.
\end{equation}
\footnotetext{This picture is taken from \cite{hobson}.}
It is also worth noting that here the parameterization parameter, $t$, of the hyper--surfaces is proportional to the proper time measured by a  comoving observer. And that the worldline for such space, given by $x^{\mu} = (t,A,B,C)$, where $A,B,\text{and }C$ are constants, satisfies the geodesic equations \eqref{geodesic}.

The metric given by equation \eqref{l1} should represent homogeneous and isotropic spacetime. Therefore, we can rewrite it as 
\begin{equation} \label{l2}
ds^2=c^2 dt^2 - S^2(t) h_{ij} dx^i dx^j. 
\end{equation}
Here, $S(t)$ is a scale factor which depends only on $t$, and $h_{ij}$ are functions of the spatial coordinates $x^i$ only. We can rewrite the spatial part, $d \sigma^2 =  h_{ij} dx^i dx^j $, in terms of spherical polar coordinates as
\begin{equation} \label{s1}
d \sigma^2 = B(r) dr^2 + r^2 (d\theta^2+ \sin^2 \theta d\phi ^2).
\end{equation}
The 3D hyper--surface should be maximally symmetric, therefore, the curvature of the space should depend only on one quantity, $K$, and for such maximally symmetric space, we can write the Ricci tensor as 
\begin{equation} \label{rms}
R_{ij} = -2 K g_{ij}.
\end{equation}  
And, if we require the metric \eqref{s1} to satisfy the condition \eqref{rms}, we find that $B(r) = 1/(1-Kr^2)$, so our spacetime metric can be rewritten as
\begin{equation} \label{l3}
ds^2=c^2 dt^2 - S^2(t) \left[ \frac{dr^2}{1-Kr^2} + r^2 (d\theta^2+ \sin^2 \theta d\phi ^2) \right].
\end{equation}
We then make the following change of variables $r \rightarrow \vert K \vert^{1/2} r $, and $k = \dfrac{K}{\vert K \vert}$, which gives
\begin{equation} \label{l4}
ds^2=c^2 dt^2 - R^2(t) \left[ \frac{dr^2}{1-kr^2} + r^2 (d\theta^2+ \sin^2 \theta d\phi ^2) \right],
\end{equation}
where $R(t)$ is given by
\[R(t) = \begin{cases}
\dfrac{S(t)}{|K|^{1/2}}  \quad \text{if } K \neq 0,
\\[0.3cm]
S(t)  \quad  \text{if }K = 0.
\end{cases}
\]
Here, $r$ is an arbitrary radial coordinate, and the coordinates $r$, $\theta$ and $\phi$ are comoving coordinates, i.e.\@  as the universe evolves in time, the worldline for any galaxy, ignoring any peculiar velocity, has fixed values of $(r,\theta,\phi)$. This is because, as we said before, the worldlines satisfy the geodesic equations.

In the metric \eqref{l4}, $k$ is $1,0 \text{ or}-1$, which corresponds to close, flat, or open universe, respectively. Investigating these cases separately, we can rewrite the metric \eqref{l4} as
\begin{equation} \label{l5}
ds^2=c^2 dt^2 - R^2(t) [ d \chi ^2 + S^2(\chi) ( d\theta^2+ \sin^2 \theta d\phi ^2) ],
\end{equation}
where $S(\chi)$ is given by
\[ S(\chi) = \begin{cases}
\sin \chi  & \text{if }k = 1, \\
\chi  & \text{if }k=0, \\
\sinh \chi & \text{if } k =-1.
\end{cases}
\]
If $k=+1$, the spatial part of equation \eqref{l5} represents a line element of a 3D hyper--surface of 4D ball with radius $R(t)$ , i.e.\@ embedded in 4D Euclidean space. If $k=0$, it represents a line element of flat 3D Euclidean space. If $k=-1$, it represents a line element of a hyperbolic 3-space embedded in 4D Minkowski space with radius $R(t)$.
\subsection{The geodesic equations in the FRW universe}

To describe particles moving due to the gravitational field of the cosmological fluid, as we did in last section, we try to find the trajectory of the freely-falling particle, i.e.\@ geodesic equations of our spacetime. We use the metric \eqref{l5}, and by applying the geodesic equation \eqref{geodesic}, we get the following conditions, i.e.\@ the geodesic equations, which hold along any geodesic in that spacetime:
\begin{equation}
\begin{array}{lllll}
\phi &=& \text{constant,} \qquad \theta &=& \text{constant,}  \\\\
R^2 \dot{\chi} &=& \text{constant,} \qquad \dot{t} &=& \begin{cases} 1 + \dfrac{R^2  \dot{\chi} ^2}{c^2} & \text{for massive particles.} \\  \dfrac{R^2  \dot{\chi} ^2}{c^2} & \text{for photons.} 
\end{cases}
\end{array}
\label{geodesicsfrw}
\end{equation}
\subsection{The cosmological redshift $z$}
In the FRW model, apart from any peculiar velocity,\footnote{Peculiar velocity in cosmology usually refers to any deviation of the galaxies velocities from that calculated from Hubble's law.}all the galaxies are comoving, i.e.\@ have fixed spatial coordinate values $(\chi,\theta,\phi)$. Therefore, the redshift in this model will be due to the expansion of the space between a stationary emitter (E), and a stationary receiver (R). Therefore, the redshift in this model can be written as
\begin{equation} \label{rsfrw}
1+z = \frac{\nu_E}{\nu_R} = \frac{p_0(R)}{p_0(E)} = \frac{R(t_R)}{R(t_E)},
\end{equation}
where $p_0(R)$, and $p_0(E)$ are the time component of the received and emitted photons' 4-momentum, respectively, and $R(t_R)$, and $R(t_E)$ are the scale factors of the universe at the reception and emission global time, respectively.
\subsection{The dynamics of the FRW spacetime}
The Einstein field equations are expressed in tensorial form as
\begin{equation} \label{efefrw}
R_{\mu \nu} - \frac{1}{2} R g_{\mu \nu} = -\kappa T_{\mu \nu} - \Lambda g_{\mu \nu} .
\end{equation}
We assume that, on a large scale, we can model all the forms of energy in the universe, i.e.\@ the cosmological fluid, as a perfect fluid. We can then make use of the energy-momentum tensor for a perfect fluid given by equation \eqref{Tperfect}. We calculate the affine connection from equation \eqref{affine} for the spacetime  that follows the line element given by equation \eqref{l5}, and calculate the Ricci tensor components from equation \eqref{riccit}. As a result, we get 2 independent equations for the scale factor,\footnote{For more detail about the derivation the reader is referred to \cite{islam}.} namely,
\begin{align}
\ddot{R}  &= -\frac{4 \pi G}{3} \left(  \rho + \frac{3p}{c^2} \right) R + \frac{1}{3} \Lambda c^2 R, \label{frwdynamics1} \\
\dot{R}^2 &= \frac{8 \pi G}{3} \rho R^2 + \frac{1}{3} \Lambda c^2 R^2 - c^2 k. \label{frwdynamics2}
\end{align}
Here, $\rho$ stands for the total energy density of matter ($m$) and radiation ($r$), so it can be written as
$\rho=\rho_{_{m}} +\rho_{_{r}}$. The two equations \eqref{frwdynamics1} and \eqref{frwdynamics2} are dubbed the Friedmann--Lema\^itre equations, and if $\Lambda =0$, they are dubbed the Friedmann equations.

If we define the dark energy density $\rho_{_{\Lambda}} $ such that $ \Lambda = \dfrac{8 \pi G}{c^2} \rho_{_{\Lambda}} $, where $\Lambda$ is the cosmological constant, and $\rho_{_{\text{total}}}$ as $\rho_{_{\text{total}}} = \rho_{_{\Lambda}} +\rho =\rho_{_{\Lambda}} +\rho_{_{r}} +\rho_{_{m}} ,$ then we can rewrite equations \eqref{frwdynamics1} and \eqref{frwdynamics2} as 
\begin{align}
\ddot{R}& = -\frac{4 \pi G}{3} \left(  \rho + \frac{3p}{c^2}  - 2 \rho_{_{\Lambda}} \right) R , \label{dyn3} \\
\dot{R}^2 &= \frac{8 \pi G}{3} \rho_{_{\text{total}}} R^2 - c^2 k = \frac{8 \pi G}{3}\left( \rho_{_{\Lambda}} +\rho_{_{r}} +\rho_{_{m}} \right) R^2 - c^2 k. \label{dyn4}
\end{align}
It is common to define the normalized scale parameter $a(t)$ as $ a(t) = \frac{R(t)}{R_0}$, where $R(t)$ is the scale factor at any time $t$ and $R_0$ is its value at certain time $t_0$, which is usually taken to be the present time. Therefore, $a_{_0} = 1$.

In terms of the normalized scale factor, we can write the energy densities as 
\begin{equation} \label{densities}
\rho_{_{\Lambda}}(t) = \rho_{_{\Lambda , 0}}, \qquad \rho_{_{m}}(t) = \dfrac{\rho_{_{m , 0}}}{a^3(t)}, \qquad     \rho_{_{r}}(t) = \dfrac{\rho_{_{r,0}}}{a^4(t)},
\end{equation}
where $ \rho_{_{\Lambda , 0}},  \rho_{_{m , 0}}$ and $ \rho_{_{r , 0}}$ refer to the present time values. The Hubble parameter is also defined as 
\begin{equation} \label{hubble}
H(t)= \dfrac{\dot{R}(t)}{R(t)} = \dfrac{\dot{a}(t)}{a(t)}.
\end{equation}
Moreover, we can define the dimensionless densities as
\begin{equation} \label{dimdensities}
\Omega_{i}(t) = \dfrac{8\pi G}{3 H^2(t)}  \rho_{_i}(t), \qquad \Omega_k(t) = -\frac{c^2 k}{H^2(t) R^2(t)},
\end{equation}
where $i$ stands for $\Lambda$, $m$, or $r$, and $ \Omega_k(t)$ is called the curvature density.

Therefore, we can rewrite equation \eqref{dyn4} as 
\begin{equation} \label{dendimdyn4}
 \Omega_{_{\Lambda}}(t) + \Omega_{m}(t) + \Omega_{r}(t) + \Omega_{k}(t) = 1,
\end{equation}
which is valid at any time. Also using equations \eqref{densities} and \eqref{dimdensities}, we can rewrite equation \eqref{dyn4} again as 
\begin{equation} \label{usedyn}
H^2(t) = H^2_{_0} \left( \Omega_{_{\Lambda,0}} + \dfrac{\Omega_{_{m,0}}}{a^3} + \dfrac{\Omega_{_{r,0}}}{a^4} + \dfrac{\Omega_{_{k,0}}}{a^2}  \right),
\end{equation}
where $H_{_0}, \Omega_{_{\Lambda,0}}, \Omega_{_{m,0}}, \Omega_{_{r,0}}$ and $\Omega_{_{k,0}}$ stand for the values at the present time.
\subsection{Friedmann Models}
The analytical solutions for equation \eqref{usedyn} with a zero-cosmological constant, i.e.\@ $\Omega_{_{\Lambda}} = 0$, are called the Friedmann models. Here, we present the solutions of the dust-dominated universes, i.e. $ \Omega_{_{r,0}} = 0.$ In this case, equation \eqref{dendimdyn4} is reduced to
\begin{equation} \label{useden}
\Omega_{m}(t) +  \Omega_{k}(t) = 1.
\end{equation}
Therefore, using equations \eqref{usedyn} and \eqref{useden}, we can write 
\begin{equation} \label{dynmat}
\dot{a}^2(t)= a^2(t) H^2(t) = H^2_{_0} \left( \dfrac{\Omega_{_{m,0}}}{a} + 1 - \Omega_{_{m,0}}  \right).
\end{equation} 
This equation can be written in integral form as 
\begin{equation} \label{tcases}
t = \frac{1}{H_{_0}} \int_0^a \left( \frac{x}{\Omega_{_{m,0}} + (1 - \Omega_{_{m,0}}) x} \right)^{1/2} dx.
\end{equation}
Here, we can separate three possible cases for the universe's geometry: flat, closed and open universes.
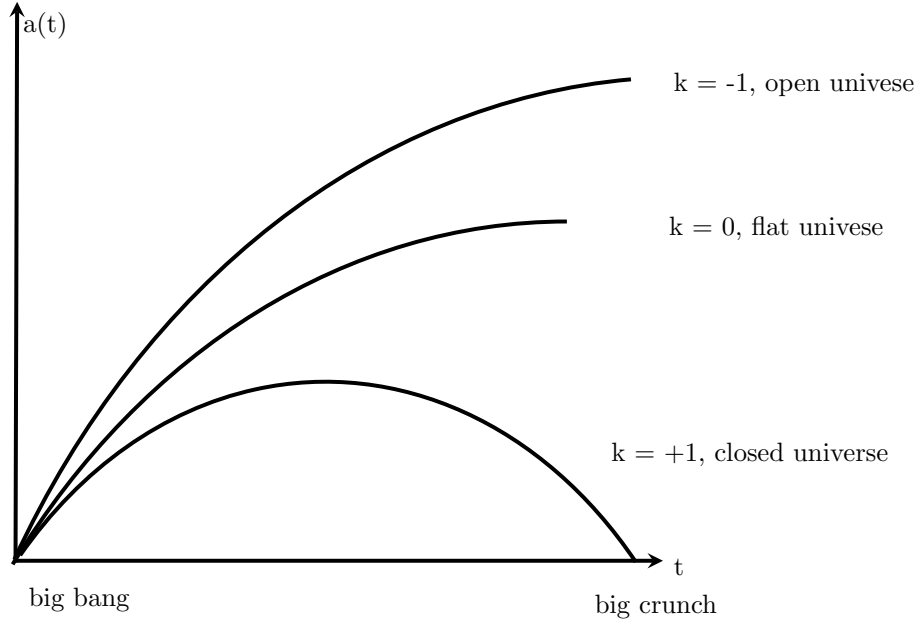
\begin{figure}[h]
\centering
\ifx\du\undefined
  \newlength{\du}
\fi
\setlength{\du}{15\unitlength}
\begin{tikzpicture}
\pgftransformxscale{0.8000000}
\pgftransformyscale{-1.000000}
\definecolor{dialinecolor}{rgb}{0.000000, 0.000000, 0.000000}
\pgfsetstrokecolor{dialinecolor}
\definecolor{dialinecolor}{rgb}{1.000000, 1.000000, 1.000000}
\pgfsetfillcolor{dialinecolor}
\pgfsetlinewidth{0.100000\du}
\pgfsetdash{}{0pt}
\pgfsetdash{}{0pt}
\pgfsetbuttcap
{
\definecolor{dialinecolor}{rgb}{0.000000, 0.000000, 0.000000}
\pgfsetfillcolor{dialinecolor}
\pgfsetarrowsend{stealth}
\definecolor{dialinecolor}{rgb}{0.000000, 0.000000, 0.000000}
\pgfsetstrokecolor{dialinecolor}
\draw (15.350000\du,14.6700000\du)--(35.550000\du,14.6700000\du);
}
\pgfsetlinewidth{0.100000\du}
\pgfsetdash{}{0pt}
\pgfsetdash{}{0pt}
\pgfsetbuttcap
{
\definecolor{dialinecolor}{rgb}{0.000000, 0.000000, 0.000000}
\pgfsetfillcolor{dialinecolor}
\pgfsetarrowsend{stealth}
\definecolor{dialinecolor}{rgb}{0.000000, 0.000000, 0.000000}
\pgfsetstrokecolor{dialinecolor}
\draw (15.350000\du,14.712500\du)--(15.400000\du,0.612500\du);
}
\pgfsetlinewidth{0.100000\du}
\pgfsetdash{}{0pt}
\pgfsetdash{}{0pt}
\pgfsetbuttcap
{
\definecolor{dialinecolor}{rgb}{0.000000, 0.000000, 0.000000}
\pgfsetfillcolor{dialinecolor}
\definecolor{dialinecolor}{rgb}{0.000000, 0.000000, 0.000000}
\pgfsetstrokecolor{dialinecolor}
\pgfpathmoveto{\pgfpoint{34.70000\du}{14.70000\du}}
\pgfpatharc{320}{221}{12.6288202\du and 12.688202\du}
\pgfusepath{stroke}
}
\pgfsetlinewidth{0.100000\du}
\pgfsetdash{}{0pt}
\pgfsetdash{}{0pt}
\pgfsetbuttcap
{
\definecolor{dialinecolor}{rgb}{0.000000, 0.000000, 0.000000}
\pgfsetfillcolor{dialinecolor}
\definecolor{dialinecolor}{rgb}{0.000000, 0.000000, 0.000000}
\pgfsetstrokecolor{dialinecolor}
\pgfpathmoveto{\pgfpoint{32.550537\du}{6.137546\du}}
\pgfpatharc{270}{217}{21.64923\du and 21.654923\du}
\pgfusepath{stroke}
}
\pgfsetlinewidth{0.100000\du}
\pgfsetdash{}{0pt}
\pgfsetdash{}{0pt}
\pgfsetbuttcap
{
\definecolor{dialinecolor}{rgb}{0.000000, 0.000000, 0.000000}
\pgfsetfillcolor{dialinecolor}
\definecolor{dialinecolor}{rgb}{0.000000, 0.000000, 0.000000}
\pgfsetstrokecolor{dialinecolor}
\pgfpathmoveto{\pgfpoint{34.550669\du}{2.562448\du}}
\pgfpatharc{266}{210}{24.161014\du and 24.271014\du}
\pgfusepath{stroke}
}
\definecolor{dialinecolor}{rgb}{0.000000, 0.000000, 0.000000}
\pgfsetstrokecolor{dialinecolor}
\node[anchor=west] at (35.550000\du,14.8150000\du){t};
\definecolor{dialinecolor}{rgb}{0.000000, 0.000000, 0.000000}
\pgfsetstrokecolor{dialinecolor}
\node[anchor=west] at (15.250000\du,1.200000\du){a(t)};
\definecolor{dialinecolor}{rgb}{0.000000, 0.000000, 0.000000}
\pgfsetstrokecolor{dialinecolor}
\node[anchor=west] at (35.550000\du,2.700000\du){k = -1, open univese};
\definecolor{dialinecolor}{rgb}{0.000000, 0.000000, 0.000000}
\pgfsetstrokecolor{dialinecolor}
\node[anchor=west] at (35.400000\du,6.300000\du){k = 0, flat univese};
\definecolor{dialinecolor}{rgb}{0.000000, 0.000000, 0.000000}
\pgfsetstrokecolor{dialinecolor}
\node[anchor=west] at (33.600000\du,12.000000\du){k = +1, closed universe};
\definecolor{dialinecolor}{rgb}{0.000000, 0.000000, 0.000000}
\pgfsetstrokecolor{dialinecolor}
\node[anchor=west] at (15.425000\du,15.645000\du){big bang};
\definecolor{dialinecolor}{rgb}{0.000000, 0.000000, 0.000000}
\pgfsetstrokecolor{dialinecolor}
\node[anchor=west] at (33.080000\du,15.805000\du){big crunch};
\end{tikzpicture}
\caption{The evolution of the scale factor in dust-dominated universes}
\label{final}
\end{figure}
\begin{itemize}
\item \textit{Flat dust-dominated universe.}
 
For such a case, we have $k = 0 $, thus $ \Omega_{_{k,0}} =0 $ and from equation \eqref{usedyn}, $ \Omega_{_{m,0}} =1 $, so equation \eqref{tcases} is reduced to
\begin{equation} \label{flat}
t = \frac{1}{H_{_0}} \int_0^a x ^{1/2} dx \Rightarrow a(t) = \left( \frac{3}{2} H_{_0} t  \right)^{\mathrm{2/3}}.
\end{equation}
This case is commonly called the Einstein--de-Sitter (EdS) model.
\newpage
\item \textit{Closed dust-dominated universe.}

For such a case, we have $k = +1$, thus $\Omega_{_{k,0}} < 0  $ and from equation \eqref{usedyn}, $ \Omega_{_{m,0}} >1 $, so we solve the integral of equation \eqref{tcases} by taking the substitution $x = \frac{\Omega_{_{m,0}}}{\Omega_{_{m,0}}-1} \sin^2 ( \psi / 2 )$, where $\psi \in [0,\pi] $ and is called the development angle. This gives
\begin{equation} \label{closed}
t = \dfrac{\Omega_{_{m,0}}}{2H_{_0}(\Omega_{_{m,0}} - 1 )^{3/2}} (\psi - \sin \psi),    \qquad a(t) = \dfrac{\Omega_{_{m,0}}}{2(\Omega_{_{m,0}} - 1 )} (1 - \cos \psi) .
\end{equation}
This shows that the graph of $a(t)$ with $t$ is cycloid, i.e.\@ $a(t)$ starts to increase until a maximum value and then starts to decrease. 
\item \textit{Open dust-dominated universe.}

For such a case, we have $k = -1 $, thus $ \Omega_{_{k,0}} > 0  $ and from equation \eqref{usedyn}, $ \Omega_{_{m,0}} < 1 $, so we solve the integral of equation \eqref{tcases} by taking the substitution $x = \frac{\Omega_{_{m,0}}}{\Omega_{_{m,0}}-1} \sinh^2 ( \psi / 2 )$, we get  
\begin{equation} \label{opened}
t = \dfrac{\Omega_{_{m,0}}}{2H_{_0}(\Omega_{_{m,0}} - 1 )^{3/2}} ( \sinh \psi - \psi),    \qquad a(t) = \dfrac{\Omega_{_{m,0}}}{2(\Omega_{_{m,0}} - 1 )} ( \cosh \psi - 1).
\end{equation}

Figure \ref{final} shows, schematically, the evolution of the scale factor $a(t)$ in the three cases discussed above.
\end{itemize}

\chapter{A Discrete Model for the Universe}

\section{Introduction}
In this chapter a model for our universe is presented. In this model the matter content is assumed to be discrete; identical\footnote{We take this case for simplicity.} spherically symmetric islands uniformly distributed in a regular lattice. This attempt was first introduced in 1957 by Lindquist and Wheeler (LW) in a seminal paper \cite{lwmodel}, where the dynamics of a closed  dust-dominated universe were studied. This model was then extended to investigate the optical properties for the flat case of such a universe in another wonderful paper by T.~Clifton and P.~Ferreira in 2009 \cite{discos}. The recent investigation showed a deviation in the optical properties for the flat case of the universe from the corresponding case in the FRW cosmology, i.e.\@ EdS model.

We start by presenting the LW model and how the dynamics for such a set-up were studied in a closed universe, and then proceed by introducing how this model was generalized in \cite{discos} to investigate the optical properties, and why this generalization is only suitable for the flat case.
\section{Lindquist Wheeler model for the universe}
This was an attempt to get the dynamics of a  dust-dominated closed universe in the discrete case. In this section we briefly cover the main points of this model; the reader is referred to \cite{lwmodel} for more details.
\subsection{The model construction}
From the FRW cosmology, as discussed below equation \eqref{l5}, the spatial part of a closed universe's line element represents the geometry of a 3D hyper-surface of a 4D ball (the comparison hyper-sphere) embedded in a 4D Euclidean space. Similarly, in the LW model, we take the universe's matter content to be discrete islands symmetrically distributed on the 3D hyper-surface of that 4D ball. As shown in \cite{regularp}, Table I(ii), it is only possible to symmetrically place $N$ points on that surface for some values of $N$, namely, \@ $N = 5,8,16,24,120,\text{or }600$. Therefore, in the LW model, our universe will simply be $N$ discrete islands (cells), each centered at one of these $N$ points (the lattice points) with point mass $m=M/N$, where $M$ is the total mass for the universe. In other words, the matter content for each cell will be modeled as a point mass placed at the center of that cell and everywhere else in the cell is empty.

On the 3D surface, a test particle is assumed to belong to an island centered at point $P$ if its closest lattice point is $P$. Between each pair of adjacent cells, there is a 2D array of points that are equally far from the centers of both cells. This array defines the 2D boundary or interface between the two cells. For each particular value of $N$, the lattice cell takes the shape of a certain curved shape of a regular polygon as show in Table \ref{table1}. Inspired by the success of the approximation methods introduced by Wigner and Seitz in solid state physics, namely, approximating regular polygons in flat space by spheres of the same volume,\footnote{For more details about this approximation method, the reader is referred to \cite{ws} Chapter 9.} Lindquist and Wheeler suggested approximating these curved polygons, i.e.\@ the lattice cells, by 3D balls. It's shown in \cite{lwmodel} that geometrically approximating the curved polygons by a sphere is more accurate than approximating the flat space cube by a sphere as assumed and showed great success in solid state physics. Otherwise stated, a curved regular polygon has closer geometric properties to a sphere of similar volume than the corresponding regular polygon in the flat space of the same volume.  
\begin{table}[here]
\centering
\begin{tabular}{c p{5cm}} 
\toprule
N & Name for the corresponding flat space polygon \\ \midrule 
5 &   tetrahedron \\
8 & cube \\ 
16 & tetrahedron \\
24 & octahedron \\  
120 & dodecahedron\\ 
600 & tetrahedron \\ \bottomrule
\end{tabular}
\caption{Closed universe cell shapes}
\label{table1}
\end{table}

The Schwarzschild solution, as we discussed, gives a line element \eqref{Sch} for the spacetime in the vicinity of a uniform distribution of material, that line element shows that the geometry of this spacetime is spherically symmetric but not uniform. The non-uniformity can be seen easily from the fact that for such line element, the Reimannian tensor $R_{ijkl}$ is not the same everywhere. However the Ricci tensor $R_{ik}$ vanishes everywhere except at the point mass where it is infinite. 

In \cite{lwmodel}, Lindquist and Wheeler replaced the spherical uniform cells on the surface of the comparison hyper-sphere by Schwarzschild cells.\footnote{A Schwarzschild cell is simply a sphere but with non-uniform curvature. In other words, the geometry for such cell is the Schwarzschild geometry for a point mass $m$ centered at that lattice cell center and of the same radius as that spherical cell it replaces.} This means that we remove the uniformly curved spherical cells and replace them by non-uniform but still spherically symmetric cells.

Because of the non-zero normal derivative\footnote{In general, the potential's normal derivative measures the rate with which a test particle moves toward/away form the cell boundary.} of the Schwarzschild potential on the cell boundaries, the masses from the two sides of the boundary will accelerate, in general, differently to nullify the difference in the normal derivative of the potential on the two sides. This leads to a relative motion between the lattice boundary and the lattice central mass, which is the main reason for this model to have dynamics.
\subsection{The mathematical description for the model}
We choose the Schwarzschild cells with total volume equal exactly to the total volume for the 3D surface of a 4D ball, i.e.\@ $2\pi^2 R^3$, where $R$ is the radius for that 4D ball. This can be mathematically written as
\begin{equation} \label{vol}
\dfrac{1}{N} = \dfrac{2 \psi_{_N} - \sin 2 \psi_{_N}}{2 \pi},
\end{equation}
where $\psi_{_N}$ is the angular separation between the two lines connecting the center of a lattice cell and its boundary to the center of the 4D ball.

The radius of the lattice cell, $a$, and the radius of the 4D ball, $R$, are related by 
\begin{equation} \label{radii}
\sin \psi_{_N} = \dfrac{a}{R}.
\end{equation}

The other necessary condition is that the new lattice cells (the patches) should be tangent to the surface of the comparison hyper-sphere. We can formulate that as follows; if we imagine making a measurement for a great circle circumference on a lattice cell, i.e.\@ we make that measurement on a great circle on the cell boundary, and the circumference of infinitesimally smaller circle. Then the following condition should be satisfied:
\begin{equation} \label{cir}
\dfrac{1}{2\pi} \dfrac{d(\text{circumference})}{d(\text{radial distance})} = \cos \psi_{_N},
\end{equation}
where $\psi_{_N}$ is as defined in equation \eqref{vol}.
\begin{figure} [h!]
\centering
\includegraphics[scale=0.5]{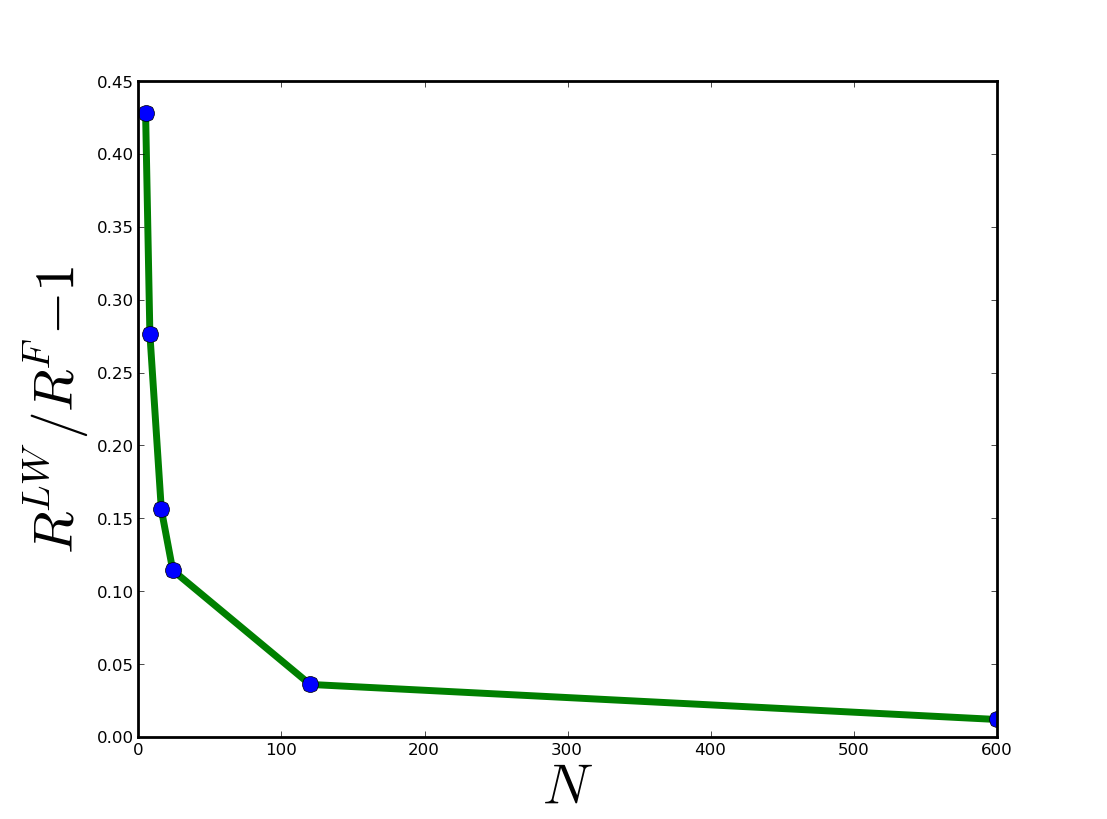}
\caption{The dependence of The ratio between the maximum radii for the comparison hyper-sphere in Lindquist--Wheeler universe, $R^{^{\text{LW}}}$, and Friedmann universe, $R^{^{\text{F}}}$, on the number of the cells $N$ for the closed universe.}
\label{figure8}
\end{figure}

To find the dynamics in the model, Lindquist and Wheeler replaced Schwarzschild time coordinate, $t$, and Schwarzschild radial distance, $r$, in the vicinity of the boundary by a new set of coordinates $\tau(r,t)$ and $ \rho(r,t)$. They constructed $ \rho(r,t)$ such that the boundary at any time is defined by $ \rho(r,t) = $ constant, and then they constructed $\tau$  to be perpendicular to $\rho $. By formulating the conditions \eqref{vol} and \eqref{cir} in terms of the new coordinates, in \cite{lwmodel} it was proven that the dynamics of the lattice cells can be described in terms of the comparison hyper-sphere radius, $R$, as
\begin{equation} \label{dy1}
\dfrac{\dot{R}^2}{R^2} = \dfrac{2m}{R^3 \sin^3  \psi_{_N} } - \dfrac{1}{R^2}.
\end{equation}
Therefore, the maximum radius for such universe is given by
\begin{equation} \label{dy2}
R^{LW}_{_{\text{max}}} = \dfrac{2 M}{N \sin^3  \psi_{_N}}.
\end{equation}
Here, $M=mN$, where $M$ is the total mass of the universe. These results showed that if $N=600$ the maximum radius for the 4D ball in this model and the corresponding maximum radius in the FRW cosmology, namely, $R^{F}_{_{\text{max}}} = \frac{4M}{3\pi}$, agree to a great extent, i.e.\@ the relative difference is about $1.2 \% $. The ratio between the maximum radii\footnote{This ratio is independent of the total universe mass $M$.} is shown in Figure \ref{figure8} for the values of $N$ given in Table \ref{table1}.

The global dynamics in both cases are given by a similar evolution functional forms, and they almost agree on the maximum radius of the comparison hyper-sphere, in the limit of large $N$. Therefore, the large scale dynamics of the two cases of the universe is the same for large $N$.
\section{Extension for LW model to calculate the redshift}
As discussed in the previous section, LW developed their model for the case of closed dust-dominated universe only. They showed that this case has similar global dynamics, in the limit of a large number of discrete islands, as the solution of Friedmann equation in the case of closed dust-dominated universe, which we discussed in the previous chapter. In \cite{discos}, the main goal was to generalize such an idea for all the possible cases of the universe, namely, flat, closed and open and find a suitable global time coordinate to investigate the optical properties for such a set-up.

The Schwarzschild coordinates can be used to describe the interior of each cell very well, however, they can't be used for the whole lattice,\footnote{If the universe in closed the lattice is corresponding to the 3D surface of the comparison hyper-sphere.} because they don't match at the cells boundaries. Otherwise stated, the Schwarzschild coordinates intersect at the boundaries and do not overlap as it is required to get a smooth 3D surface on the comparison hyper-sphere surface, and as a result, at the boundaries there will be points with two different time coordinates.

As discussed in the previous section, LW developed a new time coordinate suitable to match the time coordinates between adjacent cells at the boundaries. Since we need to investigate the optical properties for such a set-up we need a time coordinate which is suitable to be used not only at the cell boundaries but also inside the cells. To do so, the proper time of a radially out-falling particle was shown, in \cite{discos}, to be suitable to be used as a global time for the whole lattice surface, which also matches at the cell boundaries, and that can be done by using the following transformation:
\begin{equation} \label{trans}
d\tau = \sqrt{E} dt - \dfrac{\sqrt{E-1+2m/r}}{1-2m/r} dr.
\end{equation}
Here, $\tau$ is the proper time along the trajectory of radially out-falling particles, $r \text{ and } t$ are the Schwarzschild coordinates and $E$ is a constant $>0$. If we substitute the transformation \eqref{trans} in the Schwarzschild line element \eqref{Sch}\footnote{We take system of units such that $c^2=1.$}, we can write
\begin{equation} \label{lelement}
ds^2 = \frac{1}{E} \left( 1- \dfrac{2m}{r} \right) d\tau^2 + \dfrac{2}{E} \sqrt{E-1+\dfrac{2m}{r}} dr d\tau - \frac{ dr^2 }{E}- r^2 d \theta ^2 - r^2 \sin^2 \theta d \phi ^2.
\end{equation} 
Along the radially out-falling particle trajectory, $ds^2 = d\tau^2$ and $d\theta = d\phi = 0$, which gives
\begin{equation} \label{radial}
\left( \dfrac{dr}{d\tau} \right) ^2 = E -1 + \dfrac{2m}{r}. 
\end{equation}
Therefore, the $4$-velocity along such trajectory is given by\footnote{We taking the positive square root because we consider the out-falling particle.}
\begin{equation} \label{ue}
u^i = \left( 1,\sqrt{E -1 + \dfrac{2m}{r}},0,0  \right).
\end{equation}
Using the metric given by equation \eqref{lelement}, we can calculate the covariant components of the 4-velocity given by equation \eqref{ue} to find that $u_i = (-1,0,0,0)$. Therefore, we can easily prove that the surfaces of constant $\tau$ are perpendicular to the trajectory of the radially out-falling particles, by taking any arbitrary 4-vector on a surface of constant $\tau$, which is given by 
\begin{equation} \label{sur}
n^i = (0,n^r,n^{\theta},n^{\phi}).
\end{equation}
Which follows that
\begin{equation} \label{normalization}
n^i u_i = 0.
\end{equation}
Therefore, we can use $\tau$ as a suitable global time for the whole lattice (Universe).
\subsection{The universe construction}
Before discussing the construction, we discuss how we glue the cells of such a universe together and the geometry of these cells, we should also specify which regular polygon they will correspond to; in the two cases of open and flat universe the possible shapes are given in Table \ref{table2} and for the closed universe case the cells are curved and the corresponding flat cell shapes are given in Table \ref{table1}. This problem was discussed in details by H.~Coxeter in \cite{regularp}.
\begin{table} [here]
\centering
\begin{tabular}{cc p{5cm}} 
\toprule
Universe & N & Name for the corresponding flat space polygon \\ \midrule 
flat & $ \infty $ &   cube \\
open & $ \infty $& cube \\ 
open& $ \infty $ & dodecahedron\\ 
open & $ \infty $ & Icosahedron \\ \bottomrule
\end{tabular}
\caption{Flat and Open universe cell shapes}
\label{table2}
\end{table}

To construct such a set-up, we follow the following steps:
\begin{enumerate}
\item We choose one of the regular lattices with cell shapes given in Table \ref{table1} or \ref{table2}.
\item At a certain initial time we place a mass $m$ at the center of each cell.
\item Replace each regular polygon by a sphere, and as a consequence, the lattice will have overlap and no man's land regions. Figure \ref{zones} shows how we can do that for a 2-dimensional flat universe.\footnote{It was indicated in \cite{lwmodel} that there will be no region common to three cells if we construct them under the condition \eqref{vol}.}
\begin{figure}[h]
\centering
\includegraphics[scale=0.45]{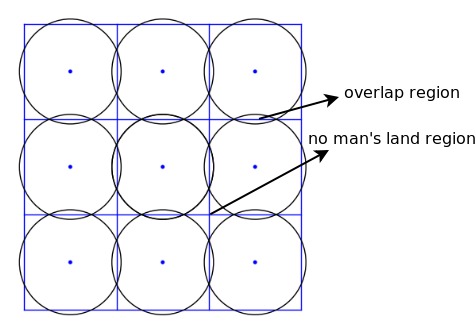}
\caption{This picture shows the overlap and no man's land regions when circles replace the squares in an 2D flat universe.}
\label{zones}
\end{figure}

\item We use the same idea used by Lindquist and Wheeler in \cite{lwmodel}, and replace the line element inside these spheres by the line element given by equation \eqref{lelement}, which is equivalent to say that we replace these uniformly curved spheres by non-uniformly curved spheres, i.e.\@ with Schwarzschild geometry. 
\item Now each sphere will have shells which are out-falling radially according to equation \eqref{radial} with the 4-velocity given by equation \eqref{ue}, so we can glue the sphere at constant $\tau$, and the orthogonality given in equation \eqref{normalization} will guarantee the tangency between these spheres and the space in which they are embedded in. However, at the overlap and the no man's land regions there will be no tangency but the tangency will exist on average as discussed in the previous section.
\end{enumerate}

We have shown how to construct the universe, and we now present the dynamics for these cells and show how, by varying the value of $E$, we get three cases for the universe, namely, flat, open and closed universes.
\subsection{The dynamics of the cells} 
According to the construction we discussed above for the universe cells (island), each spherical cell is composed of a continuous series of shells, and each shell has a particular radius, $r(\tau)$, at every global time, $\tau$. We will refer to the radius of the inner shells by $r(\tau)$, and the radius of the outermost shell (the boundary sphere) by $a(\tau)$, i.e.\@ the maximum value that $r$ can take is $a$ at the same global time, $\tau$. Therefore, we use the radius of the boundary sphere $a$ at a particular global time $\tau$ to determine the cell volume at this time, $\tau$.

Inside each cell, a particle moving with a given shell is radially out-falling according to equation \eqref{radial}. If we study the change in the boundary sphere radius, i.e.\@ replace $r(\tau)$ by $a(\tau)$ in equation \eqref{radial}, we can separate the following cases.
\begin{enumerate}
\item If $E<1$; 
the solution of equation \eqref{radial} gives that the radius of the boundary sphere increases until it reaches a maximum value at $a_{_{\text{max}}} = \dfrac{2m}{1-E}$, and begins to decrease again, i.e.\@ the cells start to collapse. This case corresponds to the closed  dust-dominated universe solution discussed in Chapter 3.
\item If $E > 1 $;
the solution of equation \eqref{radial} gives that the radius of the boundary sphere increases forever and the rate of increase when $r\rightarrow \infty $ is positive, i.e.\@ the radius will reach $\infty$ with a speed greater than zero. This case is corresponding to the open  dust-dominated universe solution discussed in Chapter 3.
\item If $E = 1 $; 
the solution of equation \eqref{radial} gives that the radius of the boundary sphere increases forever and the rate of increase when $r\rightarrow \infty $ is zero, i.e.\@ the radius will reach $\infty$ with speed equal to zero. This case is corresponding to the flat  dust-dominated universe solution (EdS model) discussed in Chapter 3.
\end{enumerate}
The three cases have the same schematic plot, shown in Figure \ref{final}.

This generalization is suitable for the flat  dust-dominated universe only. Two main reasons for saying that; first, as discussed in \cite{lwmodel}, the scale for the closed  dust-dominated universe\footnote{We mean here by the scale the maximum radius of the universe.} depends on the number of cells $N$, so if we take the flat case, the spatial part will be the Euclidean 3-space and the scale will be the same in both cases, i.e.\@ the radius reaches $\infty$ with speed zero. Second, and most important, our coordinates $(\tau,r,\theta,\phi)$ are suitable for the Euclidean space only, that is because, for instance, in closed case, they will not cover all the space, i.e.\@ in the vicinity of the maximum expansion, they will not cover all the no man's land regions. Therefore, from now on we shall take $E= 1.$ The solution for equation \eqref{radial} in the flat universe case, assuming $2m=1$ for simplicity, is $\tau-\tau_{_0} = \frac{2}{3} (r^{3/2}-r_{_0}^{3/2})$. Figure \ref{radii} shows how the radii for different shells and for the boundary sphere increase with $\tau.$ 
\begin{figure}[h!]
\centering 
\includegraphics[scale=0.7]{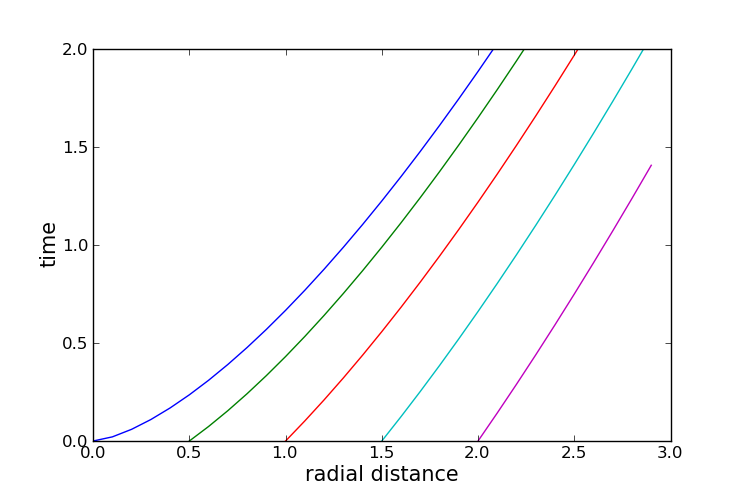}
\caption{Shows the solution for equation \eqref{radial} in the flat case,  i.e.\@ $E  = 1$. Here, we take $2m=1$ and assume that the radius of the boundary sphere is $a_{_0}= 2.0$ (the purple curve) at a certain initial time $\tau_{_0} = 0$. At $\tau = 0$,The inner shell will be smaller initial radii, and in the figure we show the change in different shells' radii with $\tau$}
\label{radii}
\end{figure}
\subsection{The geodesic equations in terms of the new coordinates} We derive the geodesic equation using the Lagrangian method, as discussed in Chapter 3. Therefore, using the line element given by equation \eqref{lelement} with $E=1$, our Lagrangian can be written as
\begin{equation} \label{lagdis}
L = g_{_{\mu \nu}} \dot{x}^{\mu} \dot{x}^{\mu} = \left( 1- \dfrac{2m}{r} \right)\dot{\tau} ^2 + 2 \sqrt{\dfrac{2m}{r}} \dot{r} \dot{\tau} - \dot{r}^2- r^2 \dot{\theta}^2 - r^2 \sin^2 \theta \, \dot{\phi} ^2. 
\end{equation}
Here, the dot refers to the derivative with respect to the parameterization parameter $\lambda$ used to parameterize this geodesic. Using the Euler--Lagrange equation $ \dfrac{d}{d\lambda} \left( \dfrac{dL}{\partial \dot{x}^{\mu}} \right)  = \dfrac{\partial L}{\partial x ^{\mu}}$, we can write that along the geodesics the following relations always hold:
\begin{equation} \label{geo1}
\frac{d}{d\lambda} \left( \dot{\tau} (1-2m/r)  +  \dot{r} \sqrt{2m/r} \right) = 0,
\end{equation}
\begin{equation} \label{geo2}
\frac{d\dot{r}}{d\lambda} - \frac{d\dot{\tau}}{d\lambda} \sqrt{2m/r} = -  \frac{m}{r^2} \dot{\tau} ^2 + r \dot{\theta}^2 + r \sin^2 \theta \, \dot{\phi} ^2,
\end{equation}
\begin{equation} \label{geo3}
\frac{d}{d\lambda} \left(r^2 \dot{\theta} \right) = r^2 \sin \theta \cos \theta \, \dot{\phi} ^2,
\end{equation}
\begin{equation} \label{geo4}
\frac{d}{d\lambda} \left( r^2 \sin^2 \theta\dot{\phi} \right)= 0.
\end{equation}
Equations \eqref{geo1} and \eqref{geo4} give
\begin{equation} \label{geo5}
\dot{\tau} (1-2m/r)  +  \dot{r} \sqrt{2m/r} = B,
\end{equation}
\begin{equation} \label{geo6}
\dot{\phi}= \dfrac{J_{\phi}}{r^2 \sin^2 \theta},
\end{equation}
where $B$ and $J_{\phi}$ are constants. Equation \eqref{geo3} can be rewritten as 
\begin{equation} \label{geo7}
\ddot{\theta} +  2 \dfrac{\dot{r}}{r} \dot{\theta} - \frac{J_{\phi}^2 \cos \theta }{r^4 \sin^3 \theta} = 0.
\end{equation}
If we multiply equation \eqref{geo7} by $r^4 \dot{\theta} $, we can write $\dfrac{d}{d\lambda} \left(  r^4 \dot{\theta}^2 +  \dfrac{J_{\phi}^2}{\sin^2 \theta} \right) = 0$. It follows that 
\begin{equation} \label{jj}
 r^4 \dot{\theta}^2 + \dfrac{J_{\phi}^2}{\sin^2 \theta} = J^2,
\end{equation}
where $J$ is a constant. Therefore, we can write that
\begin{equation} \label{geo8}
\dot{\theta}^2 = \dfrac{J^2}{ r^4} -  \frac{J_{\phi}^2}{r^4 \sin^2 \theta}.
\end{equation}
The 4-velocity for a particle moving geodesically is determined by the functions $\dot{\tau} , \dot{r}, \dot{\theta},\text{and } \dot{\phi}$ which obey equations \eqref{geo1} -- \eqref{geo4}, equations \eqref{geo8} and \eqref{geo6} determine $\dot{\theta}$ and $\dot{\phi} $, respectively. Now, as we did before, we can separate the two case of photons and massive particles.
\begin{enumerate}
\item Photons:
To get the full photons trajectory, instead of using equation \eqref{geo2} and \eqref{geo5} to determine $\dot{\tau}$ and $\dot{r}$, we use the null-constraint $g_{_{\mu \nu}} \dot{x}^{\mu} \dot{x}^{\mu} = 0$ with equation \eqref{geo5}. We can easily show that 
\begin{equation} \label{geo9}
\dot{r}^2  = B^2 - \dfrac{J^2}{ r^2} \left( 1-\dfrac{2m}{r} \right).
\end{equation}
Therefore, equations \eqref{geo5}, \eqref{geo6}, \eqref{geo8} and \eqref{geo9} determine the geodesic of a photon in the cell.   
\item Massive particles: Similarly we use the null-constraint $g_{_{\mu \nu}} \dot{x}^{\mu} \dot{x}^{\mu} = 1$ with equation \eqref{geo5}, we can easily show that 
\begin{equation} \label{geo10}
\dot{r}^2  = \dfrac{2m}{r} - 1 + B^2 - \dfrac{J^2}{ r^2} \left( 1-\dfrac{2m}{r} \right).
\end{equation}
Therefore, equations \eqref{geo5}, \eqref{geo6}, \eqref{geo8} and \eqref{geo10} determine the geodesic of any massive particle in the cell.   

We can also prove that the trajectory of massive particles moving with the bounding spheres is a geodesic. The geodesic of a massive particle with $J = J_{\phi} = 0$, i.e.\@ $\dot{\theta} = \dot{\phi} = 0$ and $B = 1$, i.e.\@ $\dot{\tau} = 1$, so that $\dot{r} = \sqrt{2m/r}$. Which has the same $4$-velocity, given in equation \eqref{ue}, of a massive particle that is radially out-falling with a spherical shell. Therefore, such a particle is moving geodesically.\footnote{This is another reason for making this calculation for the flat universe; the particle moving radially with the shells are not moving geodesically in the other cases of the universe.}
\end{enumerate}
It is also worth noting that we did not restrict ourselves to the equatorial plane as we did in Chapter 3, because we will need to match the coordinates of adjacent cells at the boundaries, and so rotating the coordinates would make the matching process too complicated.

Now, we have a full description for the photons trajectories in each cell. To investigate the optical properties for the universe, we need to know how the photons' trajectories behave when the photons move from one cell to another. One might think that this can be easily done by making a transformation for the spatial coordinates between the two cells. However, in \cite{discos},  it was shown that this is not true, because the null--constraint in the first cell, using such transformation, will give an equation in terms of the second cell coordinates, and this equation contradicts with the null--constraint equation of the second cell.
\subsection{Matching the coordinates at the boundaries of the cells}
The main idea here is that if an observer at a cell boundary measures the photon frequency or direction, at a particular global time, using the coordinates of both cells, then, in two measurements, he should get the same results.

\begin{figure} [h!]
\centering
\includegraphics[scale=0.45]{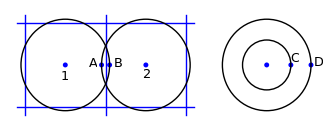}
\caption{On the left, we see two adjacent cells and two observers, A and B, on their boundary. On the right, it shows a cell with an observer on an inner shell, C, and another one on the boundary, D.}
\label{cellrmot}
\end{figure}

When photons pass from one cell to another, most of them will pass through the overlap or the no man's land regions. The photons, that will cross at intersection points, will be a set of measure zero. As shown in Figure \ref{cellrmot} on the left part, while a photon is crossing from the boundary of cell 1 to the boundary of cell 2, i.e.\@ moving between A and B, the distance A--B changes, and therefore, we expect a change in the frequency. The same thing happens if photons pass through the no man's land region, the frequencies in both cases will change in an opposite sense and on average we expect that when the photons cross a boundary, their frequency will not change. Otherwise stated, if the photon in cell 1 at the boundary has $\dot{\tau} = \dot{\tau}_1$, and at the boundary of cell 2 has $\dot{\tau} = \dot{\tau}_2$, on average, we have $\dot{\tau}_1 = \dot{\tau}_2$. This approximation has been shown to be valid numerically in \cite{discos}. 

The 4-velocity for the photons' trajectories can be decomposed as 
\begin{equation} \label{fourvel0}
k^{a} = (\dot{\tau},\dot{r},\dot{\theta},\dot{\phi}) = (-u^b k_b)(u^a + n^a),
\end{equation}
where $u^a$ is given in equation \eqref{ue} and $n^a$ is given in equation \eqref{sur}. It follows that 
\begin{equation} \label{fourvel}
k^{a} = (\dot{\tau},\dot{r},\dot{\theta},\dot{\phi}) =\dot{\tau} (1,\sqrt{2m/r} + n^r,n^{\theta},n^{\phi}) .
\end{equation}
This means that in any cell, the photons' trajectories read
\begin{equation}
\begin{array}{l}
\dot{\theta} = \dot{\tau} n^{\theta},\\
\dot{\phi} = \dot{\tau} n^{\phi},\\
\dot{r} = \dot{\tau} n^{r} + \dot{\tau} \sqrt{2m/r}.\\
\end{array} 
\label{don}
\end{equation}
Therefore, to transform a photon trajectory from one cell to another, we follow the following steps:
\begin{enumerate}
\item At the boundary of cell 1, we take $\dot{\tau}_{_1}$, $\dot{r}_{_1}$, $\dot{\theta}_{_1}$ and $\dot{\phi}_{_1}$, and use equation \eqref{don} to determine $n^{r}_{_1}$, $n^{\theta}_{_1}$ and $n^{\phi}_{_1}$.
\item Determine $n^a_{_2}$ for cell 2 from $n^a_{_2} = \dfrac{\partial x^a_{_2}}{\partial x^b_{_1}} n^b_{_1},$ where $x^a_{_1}$ and $x^a_{_2}$ are related by\footnote{Using the transformation $x_{_2} = x_{_1} - X_0$, $y_{_2} = y_{_1}$ and $z_{_2}=z_{_1}$, we find the spherical polar coordinates in cell 2 in terms of the spherical polar coordinates of cell 1. Here, $X_0$ is the cell width at the global time when the photon crosses the boundary.}
\begin{align}
r_{_2}^2 &= r^2_{_1} + X_0 ^2 - 2r_{_1}X_0 \cos \phi \sin \theta, \nonumber \\[0.3cm]
\cos ^2 \phi_{_2} &= \dfrac{(r_{_1} \cos \phi_{_1} \sin \theta_{_1} - X_0)^2}{X_0^2 + r_{_1}^2 \sin ^2 \theta_{_1} - 2r_{_1}X_0 \cos \phi_{_1} \sin \theta_{_1} }, \\[0.3cm]
\cos ^2 \theta_{_2} &= \dfrac{r_{_1}^2 \cos^2 \phi_{_1}}{X_0^2 + r_{_1}^2 \sin ^2 \theta_{_1} - 2r_{_1}X_0 \cos \phi_{_1} \sin \theta_{_1} }, \nonumber
\end{align} 
where $x^a$, or $x^b$, refers to $r$, $\theta$ and $\phi$.
\item Given that at the boundary on average $\dot{\tau}_{_1} \simeq \dot{\tau}_{_2}$, we use the results of step (2) and equation \eqref{don} to calculate $\dot{r}_{_2}$, $\dot{\theta}_{_2}$ and $\dot{\phi}_{_2}$.
\item We use these values as initial values to propagate the photons in cell 2, and upon reaching the boundary of cell 2, we take $\dot{\tau}_{_2}$, $\dot{r}_{_2}$, $\dot{\theta}_{_2}$ and $\dot{\phi}_{_2}$ and repeat the above steps.
\end{enumerate}
It is also worth noting that inside each cell the shells are out-falling according to equation \eqref{radial}, which shows that for the two observers C and D moving geodesically.\footnote{The reason for saying that they moving geodesically is discussed below equation \eqref{geo10}.} For the observers shown in the right part of Figure \ref{cellrmot}, there will be relative motion between them because they have different radial coordinate, and therefore different radial velocity, i.e.\@ there is a relative motion between C and D. This relative motion will lead to a redshift inside each cell. 

As discussed above, on average, there will be no changes in the photons' frequencies at the boundaries, which means that the redshift, in this model, occurs only inside the cells and comes only from the fact that the shells that are radially out-falling follow the 4-velocity given in equation \eqref{radial}. Now that we have the fall trajectory of the photons in the universe, i.e.\@ trajectories along which the photons travel between any two point in two different cells and across any number of cells, we can calculate the redshift occurs in this universe.

\subsection{The redshift}
We use the observers moving geodesically with the radially out-falling shells in the cell as the closest analog to the co-moving observers in an EdS model to compare the redshift in the two cases, namely, the flat matter-dominated discrete universe and the perfect fluid universe (EdS model).

The photon's frequency, emitted from, or received by, such observers, is given by $\dot{\tau} = \nu = -u^ak_{a}$, where $u^a$ and $k^{a}$ are given in equations \eqref{ue} and \eqref{fourvel}, respectively. Therefore the redshift can be expressed, in general, as 
\begin{equation} \label{disrs}
1+z = \dfrac{\nu_e}{\nu_r}=\dfrac{\dot{\tau}_e}{\dot{\tau}_r},
\end{equation}
where $\dot{\tau}_e$ and $\dot{\tau}_r$ are the values of $\dot{\tau}$ along the null-trajectory when the photon was emitted and received, respectively, and similarly for $\nu$ values. In general, tracking how $\dot{\tau}$ changes inside all the cells, the photon crossed, is only possible numerically. However, we present here an approximation to calculate it analytically in terms of the corresponding  redshift in the perfect fluid universe.
\newpage
\textbf{  $*$ An analytic approximation to calculate the redshift}

First, we try to find the redshift inside a single cell $k$, $1+\delta z_k$.

Equation \eqref{geo9}, using equations \eqref{jj} and \eqref{geo6}, can be rewritten as
\begin{equation}
\dot{r}^2  = B^2 - \dfrac{1-2m/r}{ r^2} ( r^4 \dot{\theta}^2 + r^4 \sin ^2 \theta \dot{\phi} ^2 ) = B^2 - (1-2m/r) r^2 ( \dot{\theta}^2 + \sin ^2 \theta \dot{\phi} ^2 ).
\end{equation}
Assuming that $r \gg 2m$, we can write that $B = \sqrt{\dot{r}^2 + r^2 ( \dot{\theta}^2 + \sin ^2 \theta \dot{\phi} ^2 )}.$ Therefore, in this approximation, $B$ expresses the total velocity of the photons inside each cell. We will assume that it is constant for all cells.\footnote{This is consistent with the constancy in the light speed in the perfect fluid universe.}

Moreover, from equation \eqref{geo5} we can write that $\dot{\tau} = B \frac{1-\alpha \sqrt{2m/r}}{1-2m/r}$, where $\alpha = \frac{\dot{r}}{B}$, the fraction of the radial velocity ($\dot{r} =v= \sqrt{2m/r}$) with respect to the total velocity. If we take the limit that $r \gg 2m$, i.e.\@ $v$ is small, we can rewrite $\dot{\tau}$ as
\begin{equation} \label{taudot}
\dot{\tau} = B \frac{1-\alpha v}{1-v^2} \simeq B (1-\alpha v) (1+v^2) \simeq B (1-v(\alpha - v)).
\end{equation}
Now we can calculate the redshift of a given photon inside a single Schwarzschild cell. We take $\dot{\tau}_{_{\text{in}}}$ and $\dot{\tau}_{_{\text{out}}}$ to be the values of $\dot{\tau}$ when this photon enters and leaves the cell, respectively. And since the photon enters and leaves any spherical cell at the same radius,\footnote{Here, we assume that the radius of the cell will not grow much while the photon is crossing the cell.} $v_{_{\text{in}}} \simeq v_{_{\text{out}}} \simeq v \simeq \sqrt{2m/a(\tau)}$. Therefore, the redshift of a photon while it is crossing the cell $k$ can be expressed as
\begin{equation} \label{singleshift}
\begin{array}{lll}
1 + \delta z_k &=& \dfrac{ \dot{\tau}^k_{_{\text{in}}}}{\dot{\tau}^k_{_{\text{out}}}} \simeq \dfrac{1-v^k(\alpha^k_{_{\text{in}}} - v^k)}{1-v^k(\alpha^k_{_{\text{out}}} - v^k)} \\\\ &\simeq& [ 1-v^k(\alpha^k_{_{\text{in}}} - v^k) ] [1+v^k(\alpha^k_{_{\text{out}}} - v^k)] \simeq 1+ v^k(\alpha^k_{_{\text{out}}} - \alpha^k_{_{\text{in}}}).
\end{array}
\end{equation}
Second, we calculate the total redshift. To do so, we imagine that we have three points $1$, $2$ and $3$ on a photons' trajectory, and that we place observers to measure photons' frequencies, $v_1$, $v_2$ and $v_3$  at these points. The redshift between $1$ and $2$ is $1+z_{1-2} = \frac{v_1}{v_2}$. The redshift between $2$ and $3$ is $1+z_{2-3} = \frac{v_2}{v_3}$. The redshift between $1$ and $3$ is $1+z_{1-2} = \frac{v_1}{v_3} = \frac{v_1}{v_2} \frac{v_2}{v_3} = (1+z_{1-2} )(1+z_{2-3})$. Therefore, for a photon crossing $n$-cells, the total redshift is\footnote{We are assuming that the redshift inside the cells is small.}
\begin{equation} \label{prodshift}
1+z  = \prod_{k=1} ^n ( 1 + \delta z_k )\simeq 1 + \sum_{k=1}^n \delta z_k.
\end{equation} 
Here, we will make a further assumption, namely, we assume that $\dot{r}$ will not change much inside each cell, and therefore, $\dot{r}^k_{_{\text{in}}} = - \dot{r}^k_{_{\text{out}}} \Rightarrow \dot{\alpha}^k_{_{\text{in}}}  = - \dot{\alpha}^k_{_{\text{out}}} \Rightarrow \dot{\alpha}^k_{_{\text{out}}} - \dot{\alpha}^k_{_{\text{in}}} = 2 \alpha^k $. This means that, using equation \eqref{singleshift}, $\delta z_k = ~2 \alpha^k v^k = ~2 \sqrt{2m/a^k} \alpha^k$. Substituting this in equation \eqref{prodshift}, we can write that 
\begin{equation} \label{shift2}
1+z = 1+2 \sum_{k=1}^n \sqrt{\frac{2m}{a^k}} \alpha^k \simeq 1 + 2 \sqrt{\frac{2m}{a_{_0}}} \tau_{_0}^{1/3} \int \frac{dk}{\tau^{1/3}} \alpha^k.
\end{equation}
Here, as discussed in Chapter 3, we used the fact that for a flat  dust-dominated universe $a(\tau) \propto \tau^{2/3}$. As we assumed before, we ignore the expansion of the cells while a photon crosses them, so we can write 
$\frac{\Delta \tau ^k}{\Delta k} =-~2 a^k.$ This gives, $\frac{d\tau}{dk} \simeq -2a^k = -2a^k_{_0} (\tau / \tau_{_0})^{2/3}.$ If the photon crosses cell $k$ non-radially, i.e.\@ only crosses $\beta^k$ from the cell width ($2a$), we write that $\frac{d\tau}{dk} \simeq   -2a^k_{_0} \beta^k (\tau / \tau_{_0})^{2/3}.$ Substituting this in equation \eqref{shift2} gives
\begin{equation} \label{shift3}
\begin{array}{lll}
1+z &\simeq& 1 - \sqrt{\frac{2m}{a_{_0}}} \dfrac{\tau_{_0}}{a_{_0}} \int \left( \dfrac{\alpha^k}{\beta^k}\right) \dfrac{d\tau}{\tau} =  1 - \frac{ \sqrt{2m} }{a_{_0}^{2/3} } \tau_{_0} \left\langle \dfrac{\alpha^k}{\beta^k} \right\rangle \ln \left( \dfrac{\tau_e}{\tau_r} \right) \\\\
&=& 1 + \left\langle \gamma \right\rangle  \ln \left( \dfrac{a_r}{a_e} \right) 
= 1 + \left\langle \gamma \right\rangle  \ln \left( 1+z_{_{\text{FRW}}} \right) \simeq ( 1+z_{_{\text{FRW}}}  )^{  \left\langle \gamma \right\rangle }.
\end{array}
\end{equation}
Here, $ \left\langle \gamma \right\rangle = \left\langle \dfrac{\alpha^k}{\beta^k} \right\rangle $ is the ensemble average.

In \cite{discos} (Appendix E), the value of $ \left\langle \gamma \right\rangle$ was approximated geometrically to be $2/3$, however, numerically, its value was estimated, also in \cite{discos}, to be $7/10.$
\section{Discussion}
The value of $ \left\langle \gamma \right\rangle$ was estimated both analytically and numerically in \cite{discos} and was shown to be approximately of the same value. This showed that there is a deviation in the value of the redshift in the discrete case from the corresponding values in the perfect fluid case. This means that the averaging process of the spacetime, i.e.\@ taking perfect fluid assumptions does not in general commute with the evolution of the universe. Moreover, this deviation shows that we should take the effect of the inhomogeneity in our interpretation of redshift and luminosity distances.

In the Lindquist and Wheeler model and in the extension trail we discussed above, only the case of equal masses regularly distributed was considered. However, the current surveys show that the mass distribution in the universe is due to a stochastic process during the formation of the universe structure, which means that considering such a regular distribution is a huge idealization of the universe and therefore, the next step to improve this model would be taking irregular mass distributions and considering islands with different masses. This would make the calculation very difficult but as we referred above the averaging process does not commute with the universe evolution, so we expect a further deviations from the regular case considered here.

It is also worth noting that this model depends mainly on the assumption of approximating the spacetime geometry inside each discrete island by a Schwarzschild geometry which can be seen as a weak point in this model.


\chapter*{Acknowledgements}
My deepest and greatest thanks go to Almighty Allah who give me the power to do this work. I also would like to thank my family in Egypt without their incorporeal support I would not have afford to work that hard. Many thanks to Professor Bruce and Mr. Bruno for giving their guidance.


%
\renewcommand{\bibname}{References}
\nocite{*}
\bibliographystyle{plain} 
\bibliography{trial}
\addcontentsline{toc}{chapter}{References}

\end{document}